\documentclass[fleqn,10pt]{wlscirep}
\usepackage{graphicx}	% Including figure files
\usepackage{amsmath}	% Advanced maths commands
\usepackage{amssymb}	% Extra maths symbols
\usepackage{longtable}  % Long table
\usepackage{ulem}
\usepackage{ragged2e}   % justify the figure captions (\justifying in the caption) 
\usepackage{newfloat}
\usepackage{setspace}
\usepackage{xcolor}
\DeclareFloatingEnvironment[name={Extended Data Figure},fileext=lof]{extendfigure}
\DeclareFloatingEnvironment[name={Supplementary Figure},fileext=lof]{suppfigure}
\DeclareFloatingEnvironment[name={Supplementary Table},fileext=lof]{suptable}
\DeclareFloatingEnvironment[name={Extended Data Table},fileext=lof]{extendtable}
\newcommand{\Teff}{\mbox{$T_{\mathrm{eff}}$}}
\newcommand{\logg}{\mbox{$\log{g}$}}

\newcommand{\Ion}[2]{#1{\,\sc#2}}

\newcommand{\Msun}{\mbox{$\mathrm{M}_{\odot}$}}
\newcommand{\Rsun}{\mbox{$\mathrm{R}_{\odot}$}}

\title{A hot white dwarf merger remnant revealed by an ultraviolet detection of carbon}
\author[ ]{Snehalata Sahu$^{1*\dagger}$, Antoine B\'edard$^{1*\dagger}$, Boris T. G\"ansicke$^{1}$, Pier-Emmanuel Tremblay$^{1}$, Detlev Koester$^{2}$, Jay Farihi$^3$, J.J. Hermes$^{4}$, Mark A. Hollands$^{1}$,  Tim Cunningham$^{5}$, Seth Redfield$^{6}$}

\affil[1]{Department of Physics, University of Warwick, Coventry, CV4 7AL, UK}
\affil[2]{Institut für Theoretische Physik und Astrophysik, University of Kiel, 24098 Kiel, Germany}
\affil[3]{Department of Physics \& Astronomy, University College London, Gower Street, London WC1E 6BT, UK}
\affil[4]{Department of Astronomy \& Institute for Astrophysical Research, Boston University, Boston, MA 02215, USA}
\affil[5]{Center for Astrophysics, Harvard and Smithsonian, 60 Garden St., Cambridge, MA 02138, USA}
\affil[6]{Astronomy Department and Van Vleck Observatory, Wesleyan University, Middletown, CT 06459, USA}

\affil[*]{E-mail: snehalatash30@gmail.com, antoine.bedard@warwick.ac.uk}
\affil[$\dagger$]{These authors contributed equally.}

\begin{abstract}
\textbf{
Atmospheric carbon has been detected in the optical spectra of six hydrogen-rich ultra-massive white dwarfs, revealing large carbon abundances ($\log\,\mathrm{C/H}>-0.5$) attributable to the convective dredge-up of internal carbon into thin hydrogen surface layers. These rare white dwarfs likely originate from stellar mergers, making them ``smoking guns'' for one of the binary evolution channels leading to thermonuclear supernovae. However, optical spectroscopy can uncover only the most carbon-enriched objects, suggesting that many more merger remnants may masquerade as normal pure-hydrogen atmosphere white dwarfs. Here, we report the discovery of atmospheric carbon in a \textit{Hubble Space Telescope} far-ultraviolet spectrum of WD\,0525+526, a long-known hydrogen-rich ultra-massive white dwarf. The carbon abundance ($\log\,\mathrm{C/H}=-4.62$) is 4--5\,dex lower than in the six counterparts and thus detectable only at ultraviolet wavelengths. We find that the total masses of hydrogen and helium in the envelope (10$^{-13.8}$ and 10$^{-12.6}$ of the total white dwarf mass) are substantially lower than those expected from single-star evolution, implying that WD\,0525+526 is a merger remnant. Our modelling indicates that the low surface carbon abundance arises from an envelope structure in which a thin hydrogen-rich layer floats atop a semi-convection zone --- a process that has been largely overlooked in white dwarfs. Our study highlights the importance of ultraviolet spectroscopy in identifying and characterising merger remnants.}
\end{abstract}

\begin{document}
\flushbottom
\maketitle

\thispagestyle{empty}

% Introduction + Results + Discussion limited to 3,000—3,500 words.
% 6-8 display items (figures and/or tables)
% Up to 50 references (excluding those cited exclusively in Methods)

% \section*{Introduction}
% Up to 500 words and no heading

\section*{Main}
%Maybe it's worth looking at a few recent NatAst papers to see how the flow of content from the bold paragraph into the first 1-2 paragraphs of the main text looks like.
%The spacecraft \textit{Gaia} has revealed the detailed cooling track of white dwarfs in the colour-absolute magnitude diagram\cite{gaiadr2}. 

\begin{figure*}[t]
\centerline{\includegraphics[width=0.95\hsize]{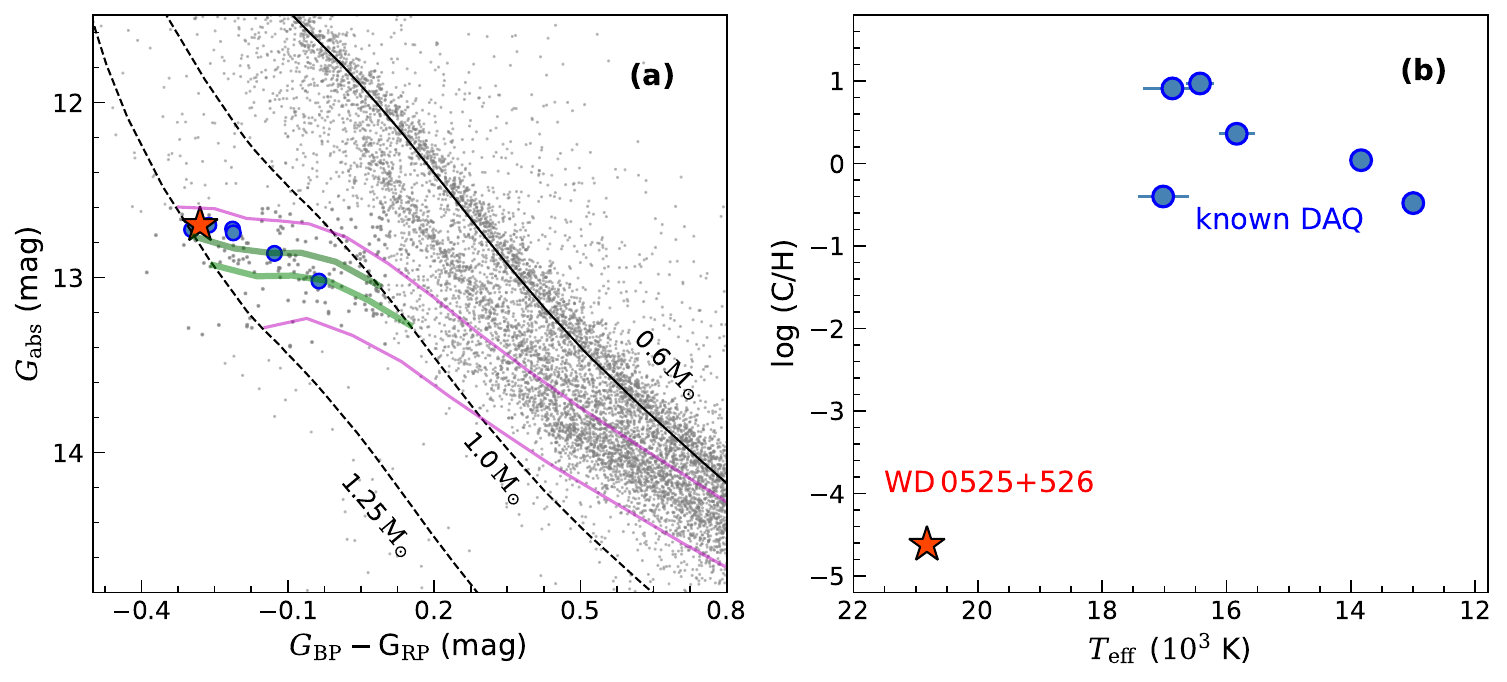}}
\caption{\justifying \textbf{DAQ white dwarfs in the \textit{Gaia} Hertzsprung-Russell diagram (HRD) and the distribution of their carbon abundances.} WD\,0525+526 is indicated by a red star, while the six published white dwarfs with spectral type DAQ \cite{mark2020, kilic2024,Jewett2024} are shown as blue circles. \textbf{(a)} \textit{Gaia} HRD showing the DAQ stars together with white dwarfs lying within 100\,pc \cite{nicola2021} (grey dots). The cooling track of a standard $0.6\,\Msun$\ hydrogen-atmosphere white dwarf\cite{Bedard2020} is over-plotted as a solid black line, and similar cooling tracks for 1.0 and $1.25\,\Msun$ white dwarfs with thinner hydrogen and helium envelope layers\cite{bedard2024} are displayed as dashed black lines. The solid magenta curves indicate where standard white dwarf models (without distillation) have a crystallised mass fraction of 20\% (top curve) and 80\% (bottom curve)\cite{Bedard2020}. The solid green curves show the beginning (top curve) and end (bottom curve) of the distillation process triggered by crystallisation in ultra-massive white dwarf models\cite{bedard2024}. We note that five DAQ stars are slightly brighter than the predicted onset of distillation; this is because the models displayed here assume a pure-hydrogen atmosphere and fixed envelope layer masses, while the DAQ white dwarfs contain atmospheric carbon and have different envelope structures. In particular, the redistribution of flux from ultraviolet to optical wavelengths by carbon lines makes white dwarfs brighter in \textit{Gaia} $G_{\rm abs}$. 
\textbf{(b)} Photospheric carbon abundances of the DAQ stars as a function of effective temperature (\Teff) with 1$\sigma$ error bars for the measurements. The carbon abundance of WD\,0525+526 is several orders of magnitude lower than those of the previously known cooler DAQ white dwarfs and could only be detected in the \textit{HST} far-ultraviolet spectrum.}
\label{fig:gaia_HRD}
\end{figure*}

The spacecraft \textit{Gaia} has revealed the detailed structure of white dwarf cooling tracks in the Hertzsprung-Russell diagram\cite{gaiadr2}, uncovering a feature known as the Q-branch, an overdensity associated with white dwarfs of all masses that experience a cooling delay of about 1\,Gyr owing to core crystallisation and related physical processes\cite{Tremblay2019} (Fig.\,\ref{fig:gaia_HRD}). The ultra-massive ($\geq1.1\,\Msun$) members of the Q-branch were found to have large space velocity dispersions indicative of abnormally old ages, implying that 5--9\% of all ultra-massive white dwarfs experience an additional cooling delay of at least 8\,Gyr\cite{Cheng2019}. The leading explanation for this extra delay is the release of gravitational energy through the solid-liquid distillation of $^{22}$Ne triggered by crystallisation in carbon-oxygen core white dwarfs\cite{simon2021, bedard2024}. This sub-population of delayed white dwarfs is interpreted as the descendants of certain types of stellar mergers, such as the merger of a white dwarf with a subgiant star \cite{shen2023}. However, it is challenging to unambiguously attribute a merger history to individual ultra-massive white dwarfs, as their core composition (carbon-oxygen or oxygen-neon) is normally obscured by layers of helium and hydrogen. Consequently, the fraction of stellar mergers on the Q-branch and the details of their past evolution remain uncertain.

The most direct observational evidence of a stellar merger history comes from the surface composition. Among ultra-massive Q-branch white dwarfs, some have apparently typical hydrogen atmospheres, but others exhibit unusually carbon-enriched atmospheres\cite{Cheng2019,coutu2019}. Among these are six white dwarfs that have mixed hydrogen-carbon (DAQ type) atmospheres with number abundance ratios $\mathrm{log(C/H)}=-0.50$ to $0.97$\cite{mark2020,kilic2024,Jewett2024}. In all six systems, the photospheric carbon has been detected in optical spectra. These rare objects have effective temperatures ranging from 13\,000 to 17\,000\,K and masses between 1.13 and 1.19\,\Msun, placing them firmly in the Q-branch (Fig.\,\ref{fig:gaia_HRD}). Their observed atmospheric compositions can only be accounted for by extremely thin hydrogen and helium layers, where underlying carbon is dredged up by a superficial convection zone\cite{Althaus2009,Koester2020}. The thin hydrogen and helium layers, coupled with the large space velocity dispersions, strongly suggest a merger origin\cite{mark2020,kilic2024}. 

\begin{figure*}[t]
\centerline{\includegraphics[width=0.95\hsize]{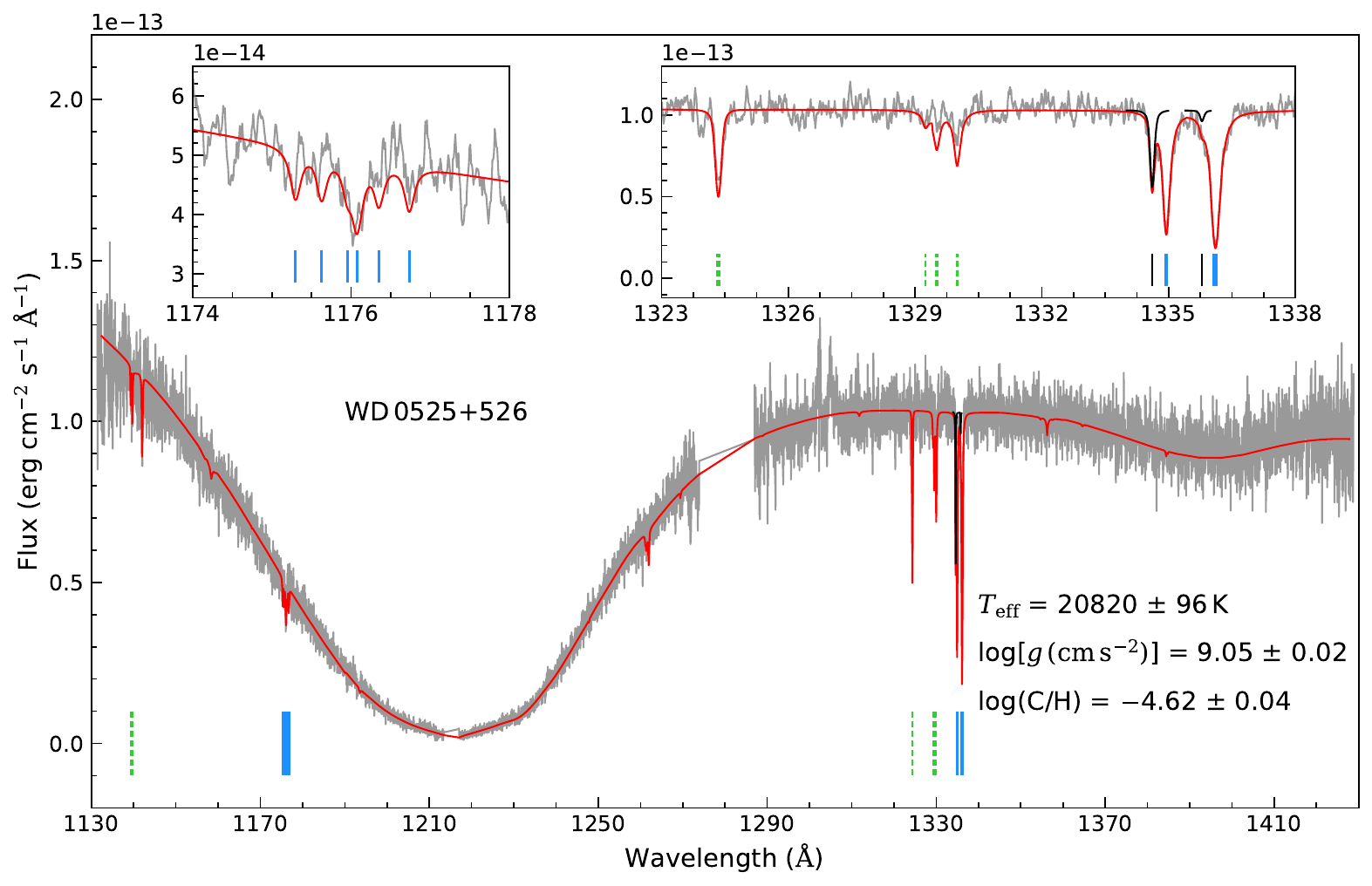}}
\caption{\justifying \textbf{\textit{HST} COS ultraviolet spectrum of WD\,0525+526.} The observed spectrum is shown in grey, and the best-fit model in red. The broad Lyman $\alpha$ absorption corroborates the hydrogen-rich atmosphere of WD\,0525+526. Multiple narrow absorption lines of carbon are detected. The stronger transitions of carbon included in the fit are indicated by solid blue lines while the weaker ones or those with inaccurate atomic data that are excluded from the fit are shown in dashed green lines. The best-fit parameters obtained from the model fit are labelled in the figure. The sub-figures provide a close-up of the (a) \Ion{C}{iii} and (b) \Ion{C}{ii} lines that were used to determine the photospheric carbon abundance. The strongest \Ion{C}{ii} lines are used to determine the radial velocity of the white dwarf which is $+92.8\pm0.6$\,km\,s$^{-1}$. The black curves show the best-fit Gaussian profiles to the carbon absorption features originating from interstellar clouds (indicated by solid black lines) traversing at an average velocity of $+18.7\pm1.1\,\mathrm{km\,s}^{-1}$ along the line of sight, which is consistent with the predicted velocity of absorption ($\approx$19.1\,km\,s$^{-1}$) from the local ISM\cite{redfield2008}.}
\label{fig:hst_spec}
\end{figure*}

Investigating the photospheric metal features present in the far-ultraviolet spectra of 311 hydrogen-atmosphere white dwarfs observed with the Cosmic Origin Spectrograph (COS) onboard the \textit{Hubble Space Telescope} (\textit{HST})\cite{sahu2023}, we detected photospheric carbon in WD\,0525+526, a white dwarf located in the high-mass end of the Q-branch (Fig.\,\ref{fig:gaia_HRD}). In addition to the broad Lyman\,$\alpha$ absorption in the spectrum confirming the atmosphere's hydrogen-dominated nature\cite{Gianninas2011}, WD\,0525+526 exhibits strong absorption lines of atomic carbon (Fig.\,\ref{fig:hst_spec}). The strongest carbon transitions are of \Ion{C}{ii} at 1334 and 1335\,\AA, which are blended with absorption arising from the interstellar medium (ISM) along the line of sight. However, the presence of higher-order atomic transitions of carbon (\Ion{C}{iii} multiplet at 1174--1177\,\AA) confirms that the red-shifted \Ion{C}{ii} lines at 92.8\,km\,s$^{-1}$ are of photospheric origin.

By fitting white dwarf atmospheric models\cite{Koester2010} to the flux-calibrated COS spectrum and the \textit{Gaia} parallax\cite{sahu2023} and using a mass-radius relation\cite{bedard2024}, we found an effective temperature of $20\,820\pm96$\,K and a mass of 1.20$\pm0.01$\,\Msun, confirming the white dwarf's ultra-massive nature. Our parameters are consistent within $\approx$3$\sigma$ with the previously reported values based on optical observations\cite{Gianninas2011, Jack2020, nicola2021}.
Keeping the white dwarf temperature and mass fixed, we modelled the carbon lines with atmospheric models
%\footnote{Note that the models are convolved with COS line spread function following the $HST$ COS documentation\href{https://spacetelescope.github.io/COS-Notebooks/LSF.html}{$^{1}$}.} 
including the contribution of the ISM lines and convolved with the COS line-spread function to derive a photospheric abundance of $\log(\rm{C/H})=-4.62\pm0.04$. The best-fitting model is shown in Fig.\,\ref{fig:hst_spec} and white dwarf parameters are provided in Table\,\ref{table:params}. We retrieved the optical spectrum of WD\,0525+526\cite{Gianninas2011} available in the Montreal White Dwarf Database \cite{dufour2017}, where neither helium nor carbon lines are detected, yielding upper limits of $\log(\mathrm{He/H})<-1.66$, and $\log\mathrm{(C/H)}<-1.91$ (both 99\% confidence; Extended\,Data\,Fig.\,\ref{fig:he_ul}).

%We retrieved the optical spectrum of WD\,0525+526\cite{Gianninas2011} available in the Montreal White Dwarf Database\cite{mwdd2017}. No He lines are detected, in particular the strong \Ion{He}{i} 4472.75\,\AA\ is not detected, and we place an upper limit of $\log(\mathrm{He/H})<-1.1$ at 99\% confidence (see Extended\,Data\,Fig.\,\ref{fig:he_ul}). We also determined an upper limit on the photospheric carbon abundance of $\log\mathrm{(C/H)}<-1.7$ based on the optical spectrum of WD\,0525+526 alone.

We conclude that WD\,0525+526 joins the small class of DAQ white dwarfs\cite{kilic2024}, among which it is the hottest and nearest (Distance $=39.1\pm0.08$\,pc\cite{gaiadr3}) member. It also has the lowest carbon abundance, $\simeq4-5$\,dex lower than that of the six cooler DAQ stars (Fig.\,\ref{fig:gaia_HRD}). The detection of carbon in WD\,0525+526 was only possible because of the COS spectroscopy available for this star. Whereas metal transitions in the optical become weaker with increasing temperature, the far-ultraviolet is rich in strong metal lines\cite{boris2012}, allowing the detection of small amounts of photospheric metals. The identification of WD\,0525+526 as a DAQ among $\simeq20$ white dwarfs with masses $>$1.1\,\Msun\ in the $\simeq1000$-strong spectroscopically complete 40\,pc sample\cite{OBrien2024} suggests that atmospheric carbon pollution of hot ultra-massive hydrogen-rich white dwarfs may be relatively common.

The previously published DAQ white dwarfs are thought to have a convective envelope in which the elements are homogeneously mixed, indicating that the detected carbon has been dredged up from the interior\cite{mark2020, kilic2024}. In contrast, the hot hydrogen-dominated atmosphere of WD\,0525+526 is expected to be stable to convection\cite{tremblay2015}. This suggests a stratified structure where an extremely thin hydrogen-rich layer floats atop a carbon-rich envelope, the latter providing the detected atmospheric carbon through upward chemical diffusion. However, this interpretation leads to a contradiction: the carbon-rich envelope should be convective\cite{saumon2022} and the overshooting flows should efficiently mix carbon up to the surface\cite{Cunningham2019}, thereby precluding the existence of a hydrogen-rich atmosphere. 

To gain further insight into this conundrum, we modelled the distribution of the chemical elements as a function of depth in the envelope of WD\,0525+526 using the STELUM code\cite{bedard2022_stelum}. In what follows, the radial chemical profile is expressed as the elemental abundances as a function of logarithmic mass depth, $\log q = \log (1-m_{r}/M_{\rm WD})$ (where $m_{r}$ denotes the mass within radius $r$ and $M_{\rm WD}$ the total mass). We computed a static envelope model with a self-consistent chemical profile obtained by considering atomic diffusion and convective mixing in equilibrium. We assumed the atmospheric parameters inferred above, including surface abundances $\log(\mathrm{He/H})=-1.66$ and $\log(\mathrm{C/H})=-4.62$. We computed the chemical profile from the atmosphere inward, by either integrating the diffusive equilibrium equations in convectively stable regions or imposing complete mixing in convectively unstable regions\cite{Koester2020}, initially ignoring convective overshoot. Under these assumptions, the observed surface composition uniquely determines the entire envelope composition (see Methods).

\begin{figure*}[t!]
\centerline{\includegraphics[width=0.8\hsize]{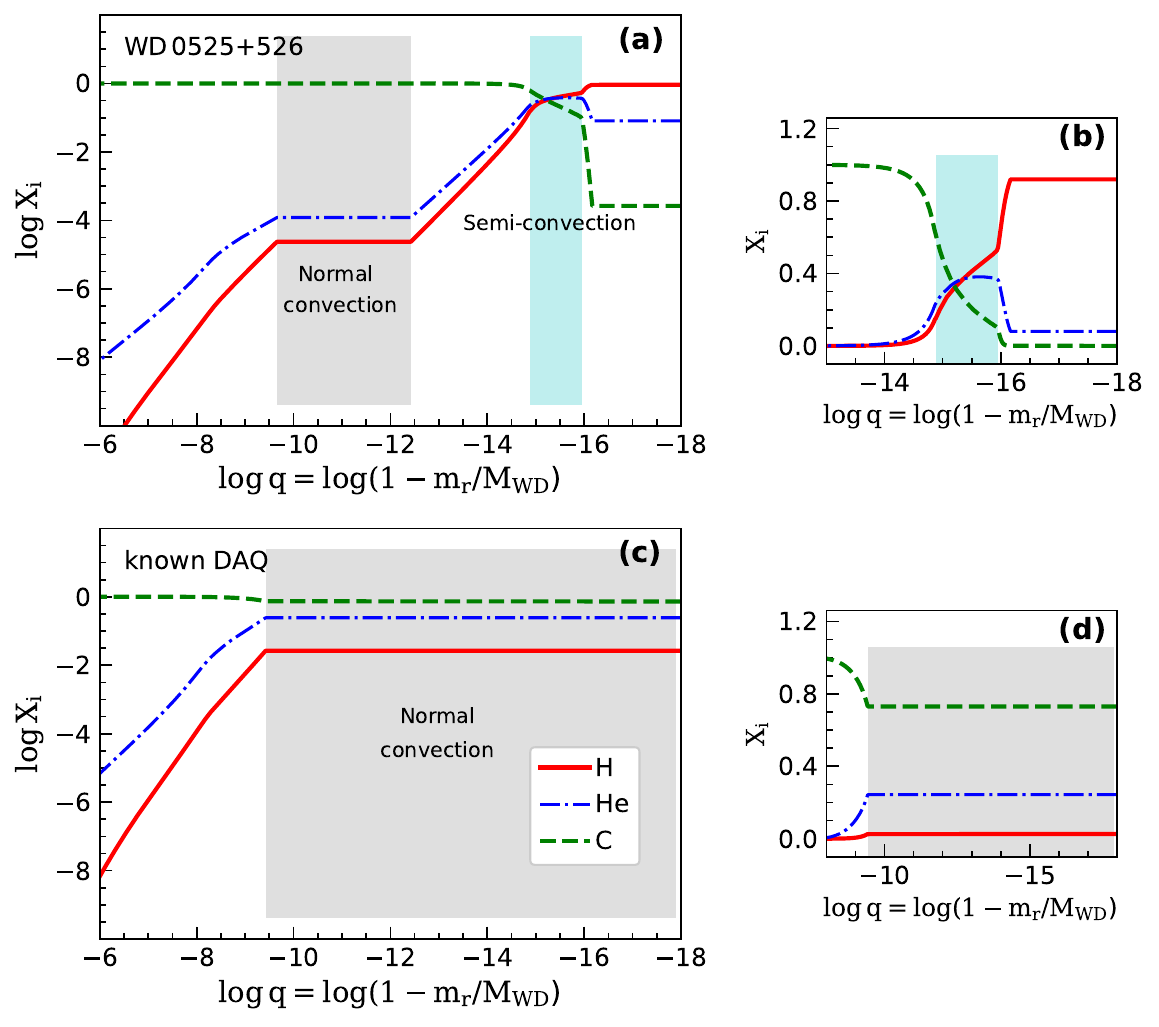}}
\caption{\justifying \textbf{Envelope chemical structure of WD$\,$0525+526 and a previously known DAQ white dwarf.} The radial chemical profile of the model envelopes is expressed as the elemental mass fractions ($X_i$ for element $i$) with logarithmic mass depth, $\log q = \log (1-m_{r}/M_{\rm WD})$. (a) Chemical profile of the hot DAQ WD$\,$0525+526 assuming the atmospheric parameters determined from ultraviolet spectroscopy (see Fig.\,\ref{fig:hst_spec}). (b) The same as (a)  but with emphasis on the outermost layers. (c) Chemical profile of the cooler DAQ WD$\,$J2340$-$1819 assuming the atmospheric parameters\cite{kilic2024} $\Teff=15\,836\,$K, $\logg=8.95$, and log(C/H)$=+0.36$, which are representative of the six known DAQ white dwarfs. (d) The same as (c)  but with emphasis on the outermost layers. In all panels, normal convection zones are shaded in grey, while semi-convection zones are shaded in blue. In the case of WD$\,$0525+526, the normal convection zone with a uniform composition is much smaller, and there is a semi-convection zone showing a composition gradient near the surface. We considered photospheric helium abundances corresponding to the upper limits determined from optical spectroscopy.}
\label{fig:env_model}
\end{figure*}

Our model (Fig.\,\ref{fig:env_model}) confirms that the atmosphere of WD\,0525+526 is not convective, allowing a thin hydrogen-helium shell to float on the surface and thus accounting for the low photospheric carbon abundance compared to the cooler DAQ stars. As anticipated, some convective mixing does occur just below the atmosphere ($\log q$ between about $-15$ and $-16$, shaded in blue in Fig.\,\ref{fig:env_model}), but we find that this is actually inefficient mixing owing to semi-convection, rather than the usual efficient mixing associated with regular convection. Semi-convection occurs when a convective instability is strongly inhibited by a composition gradient (here, the hydrogen abundance gradient), leading to partial (rather than complete) mixing\cite{salaris2017}. Although we did not include semi-convection explicitly in our modelling, it arises naturally in the form of alternating convective and non-convective layers in this region. The net result is that the slope of the abundance profiles falls in between the limiting cases of diffusive equilibrium (steeper profiles) and uniform mixing (flat profiles), which is the signature of semi-convection\cite{salaris2017} (see Methods and Extended Data Fig.\,\ref{fig:semiconv1}). The occurrence of this process in white dwarfs with extremely thin hydrogen shells was first suggested in 2007\cite{shibahashi2007, kurtz2008}, but here it is formally predicted in a detailed envelope model. 
%This result is important as semi-convection has been invoked to explain the stellar pulsations seen in the emerging class of so-called hot DAV white dwarfs\cite{kurtz2008, romero2020}. 

We note that a standard carbon-driven convection zone\cite{saumon2022} generates additional mixing in a deeper region of the envelope ($\log q$ between about $-10$ and $-12$, shaded in grey in Fig.\,\ref{fig:env_model}). As a final step, we refined our model by including convective overshoot over one pressure scale height above and below this convection zone\cite{Cunningham2019}, thereby slightly extending the mixed region (this is the final model shown in Fig.\,\ref{fig:env_model}). We did not include overshoot beyond the semi-convection zone, as appropriate for such weakened convective flows. We performed a test calculation where we imposed a uniform composition in the semi-convective region, as would be expected for efficient mixing from regular convection and overshoot, and found that the resulting model is physically inconsistent (see Methods and Extended Data Fig.\,\ref{fig:semiconv1_test}). We interpret the outcome of this test as strong evidence for inefficient mixing owing to semi-convection in WD$\,$0525+526, which effectively solves the apparent contradiction described earlier. Given that we could only determine an upper limit on the photospheric helium abundance, we also computed an envelope model without any helium, which yielded a similar chemical profile (Extended\,Data\,Fig.\,\ref{fig:env_model_nohe}).

The total mass of hydrogen contained in our model of WD\,0525+526 is $10^{-13.8}\,M_{\rm WD}$. Similarly, the upper limit on the total mass of helium (i.e., assuming the maximum surface abundance allowed by the observations) is $10^{-12.6}\,M_{\rm WD}$. Both values are many orders of magnitude lower than expected from single-star evolution\cite{renedo2010} and thus indicative of a merger history, where thermonuclear burning during and after the merger event has depleted the hydrogen and helium contents\cite{Althaus2021,shen2023}. For comparison, we also computed an envelope structure representative of the cooler and less massive DAQ stars (using the atmospheric parameters of WD\,J2340$-$1819\cite{kilic2024} as a typical example; Fig.\,\ref{fig:env_model}), which expectedly has a homogeneous outer chemical profile\cite{mark2020} ($\log q \lesssim -10$). The total hydrogen and helium masses of this model are $10^{-10.6}\,M_{\rm WD}$ and $10^{-9.3}\,M_{\rm WD}$, respectively. Therefore, WD\,0525+526 is not only hotter and more massive than other DAQ stars, but also appears to have much lower residual hydrogen and helium contents. This indicates that higher-mass mergers may lead to more severe hydrogen and helium deficiencies. 

We reviewed alternative scenarios that could account for the peculiar atmospheric composition of WD\,0525+526 and found that none of them is compelling (see Methods), leaving a merger as the most plausible explanation. Furthermore, the merger interpretation is independently supported by the location of WD\,0525+526 and other DAQ stars on the \textit{Gaia} Q-branch. Previous studies have shown that the delayed ultra-massive white dwarfs forming the Q-branch must be composed of carbon-oxygen cores\cite{Bauer2020, bedard2024}. 
Since massive single-star evolution produces white dwarfs with oxygen-neon cores\cite{Camisassa2019} (albeit with possible exceptions\cite{Althaus2021}), these objects are probably merger products\cite{Wu2022,shen2023}. 

To better constrain the merger scenario, we investigated the three-dimensional space motions of WD\,0525+526. The line-of-sight velocity determined from carbon lines in the ultraviolet spectrum is $+92.8\pm0.6$\,km\,s$^{-1}$.  Correcting for the gravitational redshift of 141.4\,km\,s$^{-1}$, based on the measured stellar mass and radius (Table~\ref{table:params}), results in a radial velocity of $-48.6\pm0.6$\,km\,s$^{-1}$.  Combined with the \textit{Gaia} proper motions, the Galactic space velocities are $(U,V,W) = (-8.1, -117.9, 1.2)$\,km\,s$^{-1}$, corrected to the local standard of rest\cite{Coskunoglu2011}. While kinematical ages are only constrained via velocity dispersions of statistical samples, the total space velocity of WD\,0525+526 is unusually high for stars in the local volume \cite{nordstrom2004}, including white dwarfs \cite{Raddi2022}, and notably lags behind Galactic rotation. Given that single stars cannot be assigned to any population with absolute confidence, the relevant probabilities (i.e. whether a given star belongs to a thin disk, thick disk or stellar halo) suggest that WD\,0525+526 is a thick disk star, where it is 280 times more likely to belong to the thick disk population than the halo population\cite{bensby2014}. This is nevertheless consistent with a star that can be as old as the disk itself, and also compatible with the other members of the DAQ spectral class \cite{kilic2024}.  Because the estimated total age under the assumption of single-star evolution is well under 1\,Gyr\cite{Bedard2020}, the space velocity supports a merger origin.

In its previous evolution, WD\,0525+526 likely exhibited an extremely hot, carbon-dominated atmosphere similar to that of the well-studied massive pre-white dwarf H1504+65\cite{werner2015}. The star then cooled down and atomic diffusion started to operate, causing residual hydrogen to float towards the surface\cite{althaus2005_H, bedard2023} and form the thin hydrogen-dominated, carbon-polluted layer that we observe today. As WD\,0525+526 cools further, it will develop efficient convection in its outer envelope, thoroughly diluting the thin hydrogen layer and dredging considerable carbon back to the surface, and will therefore turn into a DQ-type white dwarf\cite{Koester2020, kilic2024}. However, this transformation may not take place for several billion years, as the current cooling of WD\,0525+526 is likely slowed down by the energy release of the $^{22}$Ne distillation process\cite{bedard2024}. Our study highlights that only ultraviolet spectroscopy may be capable of identifying the hotter DAQ merger remnants, thereby improving our understanding of this important channel of binary evolution.
%providing important clues towards the evolution of these objects. 

\section*{Methods}

\subsection*{Optical spectroscopy}

In an attempt to improve the upper limit on the photospheric helium abundance, we acquired optical spectroscopy with the Binospec spectrograph\cite{fabricant2019-binospec} at the MMT Observatory as part of our program of spectroscopic follow-up of white dwarfs in the Solar neighbourhood (PI: T.C.). The MMT/Binospec observations were carried out in September 2024, making use of the long slit with a width of 1 arcsec. We adopted the LP3500 filter, 1000 lines/mm grating, and a central wavelength of 4,500\,\AA. This setup offers sensitivity at wavelengths ranging from 3800--5200\,\AA\ with spectral resolution $R\approx3,900$. Our observations consist of three consecutive exposures with individual exposure times of 1200\,s, yielding a total exposure time of 60 minutes. The Binospec data were reduced using the official \texttt{IDL} pipeline\cite{kansky2019}. We determined the helium and carbon upper limits following the same procedure as explained in Extended Data Fig.\ref{fig:he_ul}. The upper limits (log(He/H)$<-$1.5 and log(C/H)$<-$2.3 at 99\% confidence) were found to be consistent with those based on the optical spectrum\cite{Gianninas2011} available in the Montreal White Dwarf Database.

\subsection*{Envelope modelling}

We used the STELUM code\cite{bedard2022_stelum} to compute static envelope models for WD\,0525+526. We developed and exploited a new capability allowing to produce equilibrium models where the thermodynamic structure and chemical profile account for diffusion and convection in a self-consistent way. Such static equilibrium models are appropriate for our purpose given that the diffusive and convective timescales are both much shorter than the evolutionary timescale in the outer layers of white dwarfs\cite{fontaine1979, koester2009}. We first obtained an approximate initial model by solving the stellar structure equations for a uniform composition (i.e., imposing the measured surface abundances everywhere in the envelope). Holding the thermodynamic structure fixed, we then computed a detailed chemical profile, starting from the base of the atmosphere (Rosseland optical depth of 100, corresponding to $\log q \simeq -16.2$) and proceeding inward grid point by grid point. On convectively stable grid points, we determined the elemental abundances by integrating the diffusive equilibrium equations\cite{Koester2020}. On convectively unstable grid points, we assumed complete mixing and thus set the same composition as the grid point immediately above. We treated convection using the Schwarzschild criterion and the ML2 version of the mixing-length theory\cite{tassoul1990}. The new chemical profile was then used to update the thermodynamic structure (and hence convective region), which in turn was used to re-compute the chemical profile, with further iterations until convergence to a fully consistent model. The envelope integrations were performed down to $\log q = -3.0$, which is located well into the pure-carbon region.

We computed models for two possible atmospheric compositions, one with helium at the estimated upper limit and one without any helium. These two assumptions lead to qualitatively similar chemical profiles (shown in Fig.\,\ref{fig:env_model} and Extended\,Data\,Fig.\,\ref{fig:env_model_nohe}, respectively), and thus we focus hereafter on the model with helium. A convection zone is present at $-12 \lesssim \log q \lesssim -10$ owing to the partial K-shell ionisation of carbon\cite{saumon2022}, producing flat abundance profiles. Convection also arises closer to the atmosphere at $-16 \lesssim \log q \lesssim -15$ owing to the partial L-shell ionisation of helium and carbon\cite{cukanovaite2019, saumon2022}, but in this case the convective grid points are interspersed with non-convective grid points, resulting in non-zero abundance gradients. We interpret this behaviour as the manifestation of semi-convection\cite{salaris2017} in our simple equilibrium model.

To support our interpretation, Extended\,Data\,Fig.\,\ref{fig:semiconv1} displays the radial profiles of key physical quantities in the region of interest; the panels show, from top to bottom, the hydrogen mass fraction $X_{\rm H}$, the difference between the radiative and adiabatic temperature gradients, $\nabla_{\rm rad}-\nabla_{\rm ad}$, and the ratio of convective to total energy flux, $F_{\rm conv}/F_{\rm tot}$. For purely illustrative purposes, we first discuss a low-resolution model with only $\simeq$35 grid points in the range $-16 \lesssim \log q \lesssim -15$, shown as blue circles. Each grid point is depicted either as a filled symbol if it is convectively stable ($\nabla_{\rm rad}-\nabla_{\rm ad} < 0$, $F_{\rm conv}/F_{\rm tot}$ = 0) or as an empty symbol if it is convectively unstable ($\nabla_{\rm rad}-\nabla_{\rm ad} > 0$, $F_{\rm conv}/F_{\rm tot}$ > 0). At $\log q \lesssim -15.9$, the envelope is stable, so the hydrogen abundance decreases inward as a result of diffusive equilibrium. The gradual enrichment in partially ionised helium and carbon (at the expense of fully ionised hydrogen) causes an increase in radiative opacity and thus in $\nabla_{\rm rad}$. At some point, the latter becomes larger than $\nabla_{\rm ad}$, implying that the plasma at this depth is convective and therefore that the composition remains constant on the next underlying grid points, given our assumption that convection generates uniform mixing. Proceeding further inward, the increase in temperature at constant composition translates into a gradual decrease in opacity and $\nabla_{\rm rad}$ (a well-known feature of stellar envelopes). The latter eventually drops back below $\nabla_{\rm ad}$, so the corresponding grid point is convectively stable and the composition is once again allowed to vary following diffusive equilibrium. However, the decrease in hydrogen abundance over the next grid point immediately restores the condition of convective instability and thus chemical mixing. This pattern of alternating convective and non-convective layers repeats itself until $\log q \simeq -14.9$, where the hydrogen content becomes too low to notably affect the opacity. From this point inwards, $\nabla_{\rm rad}$ monotonically decreases and thus the plasma remains stable.

The net result of these oscillations in $\nabla_{\rm rad}-\nabla_{\rm ad}$ around 0 is a staircase-like chemical profile, with small discontinuities at non-convective grid points. These discontinuities are likely not physical: they are due to the finite number of grid points in our model as well as our assumption of complete mixing in convective regions. The latter hypothesis is probably inappropriate here given that the plasma is just on the edge between stability and instability. The same can be said of the blind use of the mixing-length formalism to calculate the convective flux and velocity, which abruptly vanish at non-convective grid points given the strictly local nature of the theory. Still, the emergence of a staircase-like chemical profile is informative as it constitutes a well-known signature of semi-convection in one-dimensional stellar models\cite{grossman1996, paxton2018} and even in three-dimensional hydrodynamical simulations\cite{wood2013}. To better capture the true chemical profile, we computed models with higher grid resolutions and indeed found that the discontinuities become smaller as the number of grid points is increased. A model with $\simeq$100 grid points in the region of interest, a threefold increase in resolution compared to the illustrative model discussed above, is displayed as red diamonds in Extended\,Data\,Fig.\,\ref{fig:semiconv1}. To remove the remaining unphysical discontinuities, the abundance profiles of this model were smoothed by imposing strict convective neutrality, $\nabla_{\rm rad}$ = $\nabla_{\rm ad}$, resulting in the black lines in Extended\,Data\,Fig.\,\ref{fig:semiconv1}. This corresponds to our adopted chemical profile shown in Fig.\,\ref{fig:env_model} (and similarly in Extended\,Data\,Fig.\,\ref{fig:env_model_nohe} for the helium-free case). The outcome is a mild, continuous composition gradient in this region, intermediate between the much steeper gradient expected from pure diffusive equilibrium and the perfectly flat gradient expected from efficient convective mixing.

To further demonstrate that this is the sole physically viable solution, we performed numerical experiments where we forced the chemical profile in the region of interest to follow either one of the two limiting cases. These artificial models are shown in Extended\,Data\,Fig.\,\ref{fig:semiconv1_test}, which is identical in format to Extended\,Data\,Fig.\,\ref{fig:semiconv1}; the blue circles and red diamonds denote the cases of imposed diffusive equilibrium and uniform mixing, respectively. In the model assuming diffusive equilibrium, the hydrogen abundance sharply decreases inward, such that $\nabla_{\rm rad}$ becomes much larger than $\nabla_{\rm ad}$, accordingly generating a convection zone that transports over 90\% of the total energy flux. Such a vigorous convection zone should completely mix the various elements, and hence this model is clearly inconsistent. In the model assuming a uniform composition, the large hydrogen abundance causes $\nabla_{\rm rad}$ to remain smaller than $\nabla_{\rm ad}$ and thus entirely suppresses the convective instability, again leading to a contradiction. Therefore, the only self-consistent solution is an intermediate composition gradient such that the equality $\nabla_{\rm rad}$ = $\nabla_{\rm ad}$ is strictly satisfied, as in our adopted model. The physical interpretation is that this region experiences semi-convection: weak convective flows inducing just enough mixing to maintain convective neutrality\cite{salaris2017}. This is a self-regulating process, as any deviation towards stronger or weaker mixing would drive the system back to equilibrium, thus leading to a unique chemical profile.

Finally, we note that our test where uniform mixing was imposed effectively rules out the possibility of mixing due to overshoot beyond the semi-convective zone. Indeed, adding overshoot in our model would shift the boundary of the mixed region closer to the surface and thus increase the hydrogen abundance in that region. This would in turn push the model farther from the condition of convective instability, worsening the contradiction described above. Therefore, we conclude that overshoot is negligible in this case, in line with the idea that fluid motions arising from semi-convection are considerably weaker than those characterising normal convection. This justifies our choice to include overshoot solely beyond the standard convection zone at $-12 \lesssim \log q \lesssim -10$.

In our envelope calculations, we placed the outer boundary at Rosseland optical depth of 100 corresponding to $\log q \simeq -16.2$, implying that we imposed a uniform composition above that point, as can be seen in Fig.\,\ref{fig:env_model}. We made this choice for consistency with the atmosphere models used in the spectroscopic analysis, which assume a homogeneous chemical mixture\cite{Koester2010}. In reality, since the atmosphere is radiative, diffusion is expected to operate all the way out to the surface. To verify that the presence of the semi-convection zone is robust to this assumption, we also computed envelope models with an outer boundary at lower optical depth. We found that such models still contain a semi-convective region of similar extent, the only difference being that this region is slightly shifted upwards (approximately following $\Delta \log \tau \sim \Delta \log q$ where $\tau$ denotes the optical depth). Therefore, we conclude that this numerical parameter does not affect our conclusion regarding the occurrence of semi-convection in WD\,0525+526. An improved analysis would require atmosphere models that similarly take diffusion into account and thus allow a stratified chemical profile\cite{manseau2016, bedard2025}, but such models currently do not exist for hydrogen--carbon compositions and their development is beyond the scope of this paper.

\subsection*{Alternative scenarios for the atmospheric composition}

We interpreted the hydrogen-dominated, carbon-polluted atmosphere of WD\,0525+526 as the result of severe hydrogen and helium deficiencies arising from a past stellar merger. In addition to this interpretation, we explored other scenarios that could potentially account for the photospheric carbon abundance observed in WD\,0525+526.

The process of radiative levitation can support heavy elements (including carbon) at the surface of hot white dwarfs with $\Teff \gtrsim 20\,000$\,K\cite{Chayer1995}. However, this mechanism is predicted to be ineffective at the relatively low temperature and very high surface gravity of WD\,0525+526\cite{koester2014, Rouis2024}. 

Another possible source of photospheric carbon is accretion of material from the ISM\cite{Dupuis1993}, a disrupted planetary system\cite{koester2014}, or a close (sub)stellar companion\cite{wilson2021}. However, in all cases, other heavy elements would be detected alongside carbon in the ultraviolet spectrum of the white dwarf. In the case of planetary debris of standard rocky composition (such as bulk Earth), silicon would be the most noticeable element. In the case of a volatile-rich object (such as a comet), other elements like nitrogen, phosphorus, and sulphur would be observed. For accretion from the ISM or a companion, all these species may be present.
%For example, a few DA white dwarfs with similar effective temperature as WD\,0525+526 are known with carbon abundances (log(C/H)) ranging from $-$8.5 to $-$5.5\cite{koester2014}, though they have Si detection in addition to C supporting the accretion scenario. 
We do not detect photospheric silicon or any heavy element other than carbon in the COS spectrum of WD\,0525+526. 
%We determined the Si upper limit (log(Si/H)$<-8.5$) from the $HST$ COS ultraviolet spectrum of the object considering the spectral region around 1265\,\AA\ where we expect a strong Si\,II doublet feature. The C/Si ratio for WD\,0525+526 is found to be $\approx$3 orders of magnitude higher than the solar value (C/Si$\approx$10) and other DA/DB white dwarfs showing carbon in their atmospheres.
From the absence of the \Ion{Si}{ii} doublet near 1265\,\AA, we measured an upper limit $\log(\mathrm{Si/H})<-8.5$, implying a C/Si ratio $\gtrsim3$ orders of magnitude higher than typically produced by accretion\cite{koester2014}. Furthermore, we searched for infrared excess and photometric variability that would indicate the presence of a debris disk or a close companion. We built the spectral energy distribution of WD\,0525+526 using available optical and infrared photometry, including \textit{Gaia} $G$-$G_{\mathrm{BP}}$-$G_{\mathrm{RP}}$\cite{gaiadr3}, Panoramic Survey Telescope and Rapid Response System (Pan-STARRS) $grizy$\cite{panstarrs2020}, and Wide-field Infrared Survey Explorer (WISE) $W1$\cite{wright2010}. We do not detect any anomalous excess in the WISE $W1$ band. We looked for photometric variability using Zwicky Transient Facility (ZTF)\cite{Bellm2019} and Transiting Exoplanet Survey Satellite (TESS)\cite{Ricker2015} observations. We found that WD\,0525+526 is not variable on timescales of a few minutes to several hours, with variability limits of at least 0.5\% in ZTF DR20 and 0.4\% from about three months (Sectors 19, 59, and 73) of 2-min-cadence data in TESS. Therefore, all variants of the accretion scenario seem highly unlikely.

Many cool white dwarfs with $\Teff \lesssim 10\,000$ K exhibit a helium-dominated, carbon-polluted atmosphere\cite{coutu2019, blouin2023} ascribed to dredge-up of internal carbon by a helium-driven convection zone\cite{althaus2005, bedard2022}. These objects are believed to originate from single-star evolution involving a so-called late helium-shell flash, a phenomenon that drastically reduces the hydrogen content while leaving the helium content largely intact\cite{althaus2005, werner2006}. The carbon dredge-up process requires a deep convection zone and thus usually occurs at effective temperatures lower than that of WD\,0525+526. There are a few exceptions, notably a handful of helium-rich white dwarfs having temperatures ($\Teff \simeq 22\,000-26\,000$\,K) and carbon abundances ($\log(\mathrm{C/H}) \simeq -5.5$) similar to those of WD\,0525+526 (while being devoid of other heavy elements)\cite{Petit2005, koester2014a}. 
%However, unlike the case of the hot DAQ, dredge-up of carbon from the thick He convection zones has been suggested as the most likely explanation for their high carbon abundances\cite{koester2014a}. In addition, most of the stars with detected radiative levitation or dredge-up have lower masses, ($0.6\,\Msun$ on average) compared to the ultra-massive nature of WD\,0525+526. 
The cause of the carbon pollution in these stars is still debated but has been tentatively attributed to convective dredge-up\cite{koester2014a}. In any case, these objects are notably different from the DAQ white dwarfs, as they have typical masses ($\approx 0.6\,\Msun$ on average) and helium-dominated atmospheres, and therefore they likely have a distinct origin\cite{bedard2024_review}.

\section*{Data availability}
The COS spectra of WD\,0525+526 is publicly available in MAST $HST$ archive (\url{https://mast.stsci.edu/search/ui/#/hst}) under program ID 15073. 
The MMT/Binospec spectrum, best-fitting model spectrum and chemical profile are provided in Supplementary Data 1-6.

\section*{Code availability}
The Koester model atmosphere code\cite{Koester2010} and the STELUM evolution code\cite{bedard2022_stelum} are not made available as they require substantial expert training and the authors do not have the resources to support general public use. However, their associated references can be consulted for further details.

\bibliography{ref}

\begin{thebibliography}{10}
\urlstyle{rm}
\expandafter\ifx\csname url\endcsname\relax
  \def\url#1{\texttt{#1}}\fi
\expandafter\ifx\csname urlprefix\endcsname\relax\def\urlprefix{URL }\fi
\expandafter\ifx\csname doiprefix\endcsname\relax\def\doiprefix{DOI: }\fi
\providecommand{\bibinfo}[2]{#2}
\providecommand{\eprint}[2][]{\url{#2}}

\bibitem{gaiadr2}
\bibinfo{author}{{Gaia Collaboration}} et~al.
\newblock \bibinfo{journal}{\bibinfo{title}{{Gaia Data Release 2. Summary of the contents and survey properties}}}.
\newblock {\it {\JournalTitle{\aap}}} \textbf{\bibinfo{volume}{616}}, \bibinfo{pages}{A1} (\bibinfo{year}{2018}).

\bibitem{Tremblay2019}
\bibinfo{author}{{Tremblay}, P.-E.} et~al.
\newblock \bibinfo{journal}{\bibinfo{title}{{Core crystallization and pile-up in the cooling sequence of evolving white dwarfs}}}.
\newblock {\it {\JournalTitle{\nat}}} \textbf{\bibinfo{volume}{565}}, \bibinfo{pages}{202--205} (\bibinfo{year}{2019}).

\bibitem{Cheng2019}
\bibinfo{author}{{Cheng}, S.}, \bibinfo{author}{{Cummings}, J.~D.} \& \bibinfo{author}{{M{\'e}nard}, B.}
\newblock \bibinfo{journal}{\bibinfo{title}{{A Cooling Anomaly of High-mass White Dwarfs}}}.
\newblock {\it {\JournalTitle{\apj}}} \textbf{\bibinfo{volume}{886}}, \bibinfo{pages}{100} (\bibinfo{year}{2019}).

\bibitem{simon2021}
\bibinfo{author}{{Blouin}, S.}, \bibinfo{author}{{Daligault}, J.} \& \bibinfo{author}{{Saumon}, D.}
\newblock \bibinfo{journal}{\bibinfo{title}{{$^{22}$Ne Phase Separation as a Solution to the Ultramassive White Dwarf Cooling Anomaly}}}.
\newblock {\it {\JournalTitle{\apjl}}} \textbf{\bibinfo{volume}{911}}, \bibinfo{pages}{L5} (\bibinfo{year}{2021}).

\bibitem{bedard2024}
\bibinfo{author}{{B{\'e}dard}, A.}, \bibinfo{author}{{Blouin}, S.} \& \bibinfo{author}{{Cheng}, S.}
\newblock \bibinfo{journal}{\bibinfo{title}{{Buoyant crystals halt the cooling of white dwarf stars}}}.
\newblock {\it {\JournalTitle{\nat}}} \textbf{\bibinfo{volume}{627}}, \bibinfo{pages}{286--288} (\bibinfo{year}{2024}).

\bibitem{shen2023}
\bibinfo{author}{{Shen}, K.~J.}, \bibinfo{author}{{Blouin}, S.} \& \bibinfo{author}{{Breivik}, K.}
\newblock \bibinfo{journal}{\bibinfo{title}{{The Q Branch Cooling Anomaly Can Be Explained by Mergers of White Dwarfs and Subgiant Stars}}}.
\newblock {\it {\JournalTitle{\apjl}}} \textbf{\bibinfo{volume}{955}}, \bibinfo{pages}{L33} (\bibinfo{year}{2023}).

\bibitem{coutu2019}
\bibinfo{author}{{Coutu}, S.} et~al.
\newblock \bibinfo{journal}{\bibinfo{title}{{Analysis of Helium-rich White Dwarfs Polluted by Heavy Elements in the Gaia Era}}}.
\newblock {\it {\JournalTitle{\apj}}} \textbf{\bibinfo{volume}{885}}, \bibinfo{pages}{74} (\bibinfo{year}{2019}).

\bibitem{mark2020}
\bibinfo{author}{{Hollands}, M.~A.} et~al.
\newblock \bibinfo{journal}{\bibinfo{title}{{An ultra-massive white dwarf with a mixed hydrogen-carbon atmosphere as a likely merger remnant}}}.
\newblock {\it {\JournalTitle{Nature Astronomy}}} \textbf{\bibinfo{volume}{4}}, \bibinfo{pages}{663--669} (\bibinfo{year}{2020}).

\bibitem{kilic2024}
\bibinfo{author}{{Kilic}, M.} et~al.
\newblock \bibinfo{journal}{\bibinfo{title}{{White Dwarf Merger Remnants: The DAQ Subclass}}}.
\newblock {\it {\JournalTitle{\apj}}} \textbf{\bibinfo{volume}{965}}, \bibinfo{pages}{159} (\bibinfo{year}{2024}).

\bibitem{Jewett2024}
\bibinfo{author}{{Jewett}, G.} et~al.
\newblock \bibinfo{journal}{\bibinfo{title}{{Massive White Dwarfs in the 100 pc Sample: Magnetism, Rotation, Pulsations, and the Merger Fraction}}}.
\newblock {\it {\JournalTitle{\apj}}} \textbf{\bibinfo{volume}{974}}, \bibinfo{pages}{12} (\bibinfo{year}{2024}).

\bibitem{Althaus2009}
\bibinfo{author}{{Althaus}, L.~G.}, \bibinfo{author}{{Garc{\'\i}a-Berro}, E.}, \bibinfo{author}{{C{\'o}rsico}, A.~H.}, \bibinfo{author}{{Miller Bertolami}, M.~M.} \& \bibinfo{author}{{Romero}, A.~D.}
\newblock \bibinfo{journal}{\bibinfo{title}{{On the Formation of Hot DQ White Dwarfs}}}.
\newblock {\it {\JournalTitle{\apjl}}} \textbf{\bibinfo{volume}{693}}, \bibinfo{pages}{L23--L26} (\bibinfo{year}{2009}).

\bibitem{Koester2020}
\bibinfo{author}{{Koester}, D.}, \bibinfo{author}{{Kepler}, S.~O.} \& \bibinfo{author}{{Irwin}, A.~W.}
\newblock \bibinfo{journal}{\bibinfo{title}{{New white dwarf envelope models and diffusion. Application to DQ white dwarfs}}}.
\newblock {\it {\JournalTitle{\aap}}} \textbf{\bibinfo{volume}{635}}, \bibinfo{pages}{A103} (\bibinfo{year}{2020}).

\bibitem{sahu2023}
\bibinfo{author}{{Sahu}, S.} et~al.
\newblock \bibinfo{journal}{\bibinfo{title}{{An HST COS ultraviolet spectroscopic survey of 311 DA white dwarfs - I. Fundamental parameters and comparative studies}}}.
\newblock {\it {\JournalTitle{\mnras}}} \textbf{\bibinfo{volume}{526}}, \bibinfo{pages}{5800--5823} (\bibinfo{year}{2023}).

\bibitem{Gianninas2011}
\bibinfo{author}{{Gianninas}, A.}, \bibinfo{author}{{Bergeron}, P.} \& \bibinfo{author}{{Ruiz}, M.~T.}
\newblock \bibinfo{journal}{\bibinfo{title}{{A Spectroscopic Survey and Analysis of Bright, Hydrogen-rich White Dwarfs}}}.
\newblock {\it {\JournalTitle{\apj}}} \textbf{\bibinfo{volume}{743}}, \bibinfo{pages}{138} (\bibinfo{year}{2011}).

\bibitem{Koester2010}
\bibinfo{author}{{Koester}, D.}
\newblock \bibinfo{journal}{\bibinfo{title}{{White dwarf spectra and atmosphere models}}}.
\newblock {\it {\JournalTitle{\memsai}}} \textbf{\bibinfo{volume}{81}}, \bibinfo{pages}{921--931} (\bibinfo{year}{2010}).

\bibitem{Jack2020}
\bibinfo{author}{{McCleery}, J.} et~al.
\newblock \bibinfo{journal}{\bibinfo{title}{{Gaia white dwarfs within 40 pc II: the volume-limited Northern hemisphere sample}}}.
\newblock {\it {\JournalTitle{\mnras}}} \textbf{\bibinfo{volume}{499}}, \bibinfo{pages}{1890--1908} (\bibinfo{year}{2020}).

\bibitem{nicola2021}
\bibinfo{author}{{Gentile Fusillo}, N.~P.} et~al.
\newblock \bibinfo{journal}{\bibinfo{title}{{A catalogue of white dwarfs in Gaia EDR3}}}.
\newblock {\it {\JournalTitle{\mnras}}} \textbf{\bibinfo{volume}{508}}, \bibinfo{pages}{3877--3896} (\bibinfo{year}{2021}).

\bibitem{dufour2017}
\bibinfo{author}{{Dufour}, P.} et~al.
\newblock \bibinfo{title}{{The Montreal White Dwarf Database: A Tool for the Community}}.
\newblock In \bibinfo{editor}{{Tremblay}, P.~E.}, \bibinfo{editor}{{Gaensicke}, B.} \& \bibinfo{editor}{{Marsh}, T.} (eds.) {\bibinfo{booktitle}{20th European White Dwarf Workshop}}, vol. \bibinfo{volume}{509} of {\bibinfo{series}{Astronomical Society of the Pacific Conference Series}}, \bibinfo{pages}{3} (\bibinfo{year}{2017}).

\bibitem{gaiadr3}
\bibinfo{author}{{Gaia Collaboration}} et~al.
\newblock \bibinfo{journal}{\bibinfo{title}{{Gaia Data Release 3. Summary of the content and survey properties}}}.
\newblock {\it {\JournalTitle{\aap}}} \textbf{\bibinfo{volume}{674}}, \bibinfo{pages}{A1} (\bibinfo{year}{2023}).

\bibitem{boris2012}
\bibinfo{author}{{G{\"a}nsicke}, B.~T.} et~al.
\newblock \bibinfo{journal}{\bibinfo{title}{{The chemical diversity of exo-terrestrial planetary debris around white dwarfs}}}.
\newblock {\it {\JournalTitle{\mnras}}} \textbf{\bibinfo{volume}{424}}, \bibinfo{pages}{333--347} (\bibinfo{year}{2012}).

\bibitem{OBrien2024}
\bibinfo{author}{{O'Brien}, M.~W.} et~al.
\newblock \bibinfo{journal}{\bibinfo{title}{{The 40 pc sample of white dwarfs from Gaia}}}.
\newblock {\it {\JournalTitle{\mnras}}} \textbf{\bibinfo{volume}{527}}, \bibinfo{pages}{8687--8705} (\bibinfo{year}{2024}).

\bibitem{Bedard2020}
\bibinfo{author}{{B{\'e}dard}, A.}, \bibinfo{author}{{Bergeron}, P.}, \bibinfo{author}{{Brassard}, P.} \& \bibinfo{author}{{Fontaine}, G.}
\newblock \bibinfo{journal}{\bibinfo{title}{{On the Spectral Evolution of Hot White Dwarf Stars. I. A Detailed Model Atmosphere Analysis of Hot White Dwarfs from SDSS DR12}}}.
\newblock {\it {\JournalTitle{\apj}}} \textbf{\bibinfo{volume}{901}}, \bibinfo{pages}{93} (\bibinfo{year}{2020}).

\bibitem{redfield2008}
\bibinfo{author}{{Redfield}, S.} \& \bibinfo{author}{{Linsky}, J.~L.}
\newblock \bibinfo{journal}{\bibinfo{title}{{The Structure of the Local Interstellar Medium. IV. Dynamics, Morphology, Physical Properties, and Implications of Cloud-Cloud Interactions}}}.
\newblock {\it {\JournalTitle{\apj}}} \textbf{\bibinfo{volume}{673}}, \bibinfo{pages}{283--314} (\bibinfo{year}{2008}).

\bibitem{tremblay2015}
\bibinfo{author}{{Tremblay}, P.~E.} et~al.
\newblock \bibinfo{journal}{\bibinfo{title}{{Calibration of the Mixing-length Theory for Convective White Dwarf Envelopes}}}.
\newblock {\it {\JournalTitle{\apj}}} \textbf{\bibinfo{volume}{799}}, \bibinfo{pages}{142} (\bibinfo{year}{2015}).

\bibitem{saumon2022}
\bibinfo{author}{{Saumon}, D.}, \bibinfo{author}{{Blouin}, S.} \& \bibinfo{author}{{Tremblay}, P.-E.}
\newblock \bibinfo{journal}{\bibinfo{title}{{Current challenges in the physics of white dwarf stars}}}.
\newblock {\it {\JournalTitle{\physrep}}} \textbf{\bibinfo{volume}{988}}, \bibinfo{pages}{1--63} (\bibinfo{year}{2022}).

\bibitem{Cunningham2019}
\bibinfo{author}{{Cunningham}, T.}, \bibinfo{author}{{Tremblay}, P.-E.}, \bibinfo{author}{{Freytag}, B.}, \bibinfo{author}{{Ludwig}, H.-G.} \& \bibinfo{author}{{Koester}, D.}
\newblock \bibinfo{journal}{\bibinfo{title}{{Convective overshoot and macroscopic diffusion in pure-hydrogen-atmosphere white dwarfs}}}.
\newblock {\it {\JournalTitle{\mnras}}} \textbf{\bibinfo{volume}{488}}, \bibinfo{pages}{2503--2522} (\bibinfo{year}{2019}).

\bibitem{bedard2022_stelum}
\bibinfo{author}{{B{\'e}dard}, A.}, \bibinfo{author}{{Brassard}, P.}, \bibinfo{author}{{Bergeron}, P.} \& \bibinfo{author}{{Blouin}, S.}
\newblock \bibinfo{journal}{\bibinfo{title}{{On the Spectral Evolution of Hot White Dwarf Stars. II. Time-dependent Simulations of Element Transport in Evolving White Dwarfs with STELUM}}}.
\newblock {\it {\JournalTitle{\apj}}} \textbf{\bibinfo{volume}{927}}, \bibinfo{pages}{128} (\bibinfo{year}{2022}).

\bibitem{salaris2017}
\bibinfo{author}{{Salaris}, M.} \& \bibinfo{author}{{Cassisi}, S.}
\newblock \bibinfo{journal}{\bibinfo{title}{{Chemical element transport in stellar evolution models}}}.
\newblock {\it {\JournalTitle{Royal Society Open Science}}} \textbf{\bibinfo{volume}{4}}, \bibinfo{pages}{170192} (\bibinfo{year}{2017}).

\bibitem{shibahashi2007}
\bibinfo{author}{{Shibahashi}, H.}
\newblock \bibinfo{title}{{The DB Gap and Pulsations of White Dwarfs}}.
\newblock In \bibinfo{editor}{{Stancliffe}, R.~J.}, \bibinfo{editor}{{Houdek}, G.}, \bibinfo{editor}{{Martin}, R.~G.} \& \bibinfo{editor}{{Tout}, C.~A.} (eds.) {\bibinfo{booktitle}{Unsolved Problems in Stellar Physics: A Conference in Honor of Douglas Gough}}, vol. \bibinfo{volume}{948} of {\bibinfo{series}{American Institute of Physics Conference Series}}, \bibinfo{pages}{35--42} (\bibinfo{publisher}{AIP}, \bibinfo{year}{2007}).

\bibitem{kurtz2008}
\bibinfo{author}{{Kurtz}, D.~W.}, \bibinfo{author}{{Shibahashi}, H.}, \bibinfo{author}{{Dhillon}, V.~S.}, \bibinfo{author}{{Marsh}, T.~R.} \& \bibinfo{author}{{Littlefair}, S.~P.}
\newblock \bibinfo{journal}{\bibinfo{title}{{A search for a new class of pulsating DA white dwarf stars in the DB gap}}}.
\newblock {\it {\JournalTitle{\mnras}}} \textbf{\bibinfo{volume}{389}}, \bibinfo{pages}{1771--1779} (\bibinfo{year}{2008}).

\bibitem{renedo2010}
\bibinfo{author}{{Renedo}, I.} et~al.
\newblock \bibinfo{journal}{\bibinfo{title}{{New Cooling Sequences for Old White Dwarfs}}}.
\newblock {\it {\JournalTitle{\apj}}} \textbf{\bibinfo{volume}{717}}, \bibinfo{pages}{183--195} (\bibinfo{year}{2010}).

\bibitem{Althaus2021}
\bibinfo{author}{{Althaus}, L.~G.} et~al.
\newblock \bibinfo{journal}{\bibinfo{title}{{The formation of ultra-massive carbon-oxygen core white dwarfs and their evolutionary and pulsational properties}}}.
\newblock {\it {\JournalTitle{\aap}}} \textbf{\bibinfo{volume}{646}}, \bibinfo{pages}{A30} (\bibinfo{year}{2021}).

\bibitem{Bauer2020}
\bibinfo{author}{{Bauer}, E.~B.}, \bibinfo{author}{{Schwab}, J.}, \bibinfo{author}{{Bildsten}, L.} \& \bibinfo{author}{{Cheng}, S.}
\newblock \bibinfo{journal}{\bibinfo{title}{{Multi-gigayear White Dwarf Cooling Delays from Clustering-enhanced Gravitational Sedimentation}}}.
\newblock {\it {\JournalTitle{\apj}}} \textbf{\bibinfo{volume}{902}}, \bibinfo{pages}{93} (\bibinfo{year}{2020}).

\bibitem{Camisassa2019}
\bibinfo{author}{{Camisassa}, M.~E.} et~al.
\newblock \bibinfo{journal}{\bibinfo{title}{{The evolution of ultra-massive white dwarfs}}}.
\newblock {\it {\JournalTitle{\aap}}} \textbf{\bibinfo{volume}{625}}, \bibinfo{pages}{A87} (\bibinfo{year}{2019}).

\bibitem{Wu2022}
\bibinfo{author}{{Wu}, C.}, \bibinfo{author}{{Xiong}, H.} \& \bibinfo{author}{{Wang}, X.}
\newblock \bibinfo{journal}{\bibinfo{title}{{Formation of ultra-massive carbon-oxygen white dwarfs from the merger of carbon-oxygen and helium white dwarf pairs}}}.
\newblock {\it {\JournalTitle{\mnras}}} \textbf{\bibinfo{volume}{512}}, \bibinfo{pages}{2972--2987} (\bibinfo{year}{2022}).

\bibitem{Coskunoglu2011}
\bibinfo{author}{{Co{\c{s}}kuno{\v{g}}lu}, B.} et~al.
\newblock \bibinfo{journal}{\bibinfo{title}{{Local stellar kinematics from RAVE data - I. Local standard of rest}}}.
\newblock {\it {\JournalTitle{\mnras}}} \textbf{\bibinfo{volume}{412}}, \bibinfo{pages}{1237--1245} (\bibinfo{year}{2011}).

\bibitem{nordstrom2004}
\bibinfo{author}{{Nordstr{\"o}m}, B.} et~al.
\newblock \bibinfo{journal}{\bibinfo{title}{{The Geneva-Copenhagen survey of the Solar neighbourhood. Ages, metallicities, and kinematic properties of {\ensuremath{\sim}}14 000 F and G dwarfs}}}.
\newblock {\it {\JournalTitle{\aap}}} \textbf{\bibinfo{volume}{418}}, \bibinfo{pages}{989--1019} (\bibinfo{year}{2004}).

\bibitem{Raddi2022}
\bibinfo{author}{{Raddi}, R.} et~al.
\newblock \bibinfo{journal}{\bibinfo{title}{{Kinematic properties of white dwarfs. Galactic orbital parameters and age-velocity dispersion relation}}}.
\newblock {\it {\JournalTitle{\aap}}} \textbf{\bibinfo{volume}{658}}, \bibinfo{pages}{A22} (\bibinfo{year}{2022}).

\bibitem{bensby2014}
\bibinfo{author}{{Bensby}, T.}, \bibinfo{author}{{Feltzing}, S.} \& \bibinfo{author}{{Oey}, M.~S.}
\newblock \bibinfo{journal}{\bibinfo{title}{{Exploring the Milky Way stellar disk. A detailed elemental abundance study of 714 F and G dwarf stars in the solar neighbourhood}}}.
\newblock {\it {\JournalTitle{\aap}}} \textbf{\bibinfo{volume}{562}}, \bibinfo{pages}{A71} (\bibinfo{year}{2014}).

\bibitem{werner2015}
\bibinfo{author}{{Werner}, K.} \& \bibinfo{author}{{Rauch}, T.}
\newblock \bibinfo{journal}{\bibinfo{title}{{Analysis of HST/COS spectra of the bare C-O stellar core H1504+65 and a high-velocity twin in the Galactic halo}}}.
\newblock {\it {\JournalTitle{\aap}}} \textbf{\bibinfo{volume}{584}}, \bibinfo{pages}{A19} (\bibinfo{year}{2015}).

\bibitem{althaus2005_H}
\bibinfo{author}{{Althaus}, L.~G.}, \bibinfo{author}{{Miller Bertolami}, M.~M.}, \bibinfo{author}{{C{\'o}rsico}, A.~H.}, \bibinfo{author}{{Garc{\'\i}a-Berro}, E.} \& \bibinfo{author}{{Gil-Pons}, P.}
\newblock \bibinfo{journal}{\bibinfo{title}{{The formation of DA white dwarfs with thin hydrogen envelopes}}}.
\newblock {\it {\JournalTitle{\aap}}} \textbf{\bibinfo{volume}{440}}, \bibinfo{pages}{L1--L4} (\bibinfo{year}{2005}).

\bibitem{bedard2023}
\bibinfo{author}{{B{\'e}dard}, A.}, \bibinfo{author}{{Bergeron}, P.} \& \bibinfo{author}{{Brassard}, P.}
\newblock \bibinfo{journal}{\bibinfo{title}{{On the Spectral Evolution of Hot White Dwarf Stars. IV. The Diffusion and Mixing of Residual Hydrogen in Helium-rich White Dwarfs}}}.
\newblock {\it {\JournalTitle{\apj}}} \textbf{\bibinfo{volume}{946}}, \bibinfo{pages}{24} (\bibinfo{year}{2023}).

\bibitem{fabricant2019-binospec}
\bibinfo{author}{{Fabricant}, D.} et~al.
\newblock \bibinfo{journal}{\bibinfo{title}{{Binospec: A Wide-field Imaging Spectrograph for the MMT}}}.
\newblock {\it {\JournalTitle{\pasp}}} \textbf{\bibinfo{volume}{131}}, \bibinfo{pages}{075004} (\bibinfo{year}{2019}).

\bibitem{kansky2019}
\bibinfo{author}{{Kansky}, J.} et~al.
\newblock \bibinfo{journal}{\bibinfo{title}{{Binospec Software System}}}.
\newblock {\it {\JournalTitle{\pasp}}} \textbf{\bibinfo{volume}{131}}, \bibinfo{pages}{075005} (\bibinfo{year}{2019}).

\bibitem{fontaine1979}
\bibinfo{author}{{Fontaine}, G.} \& \bibinfo{author}{{Michaud}, G.}
\newblock \bibinfo{journal}{\bibinfo{title}{{Diffusion time scales in white dwarfs.}}}
\newblock {\it {\JournalTitle{\apj}}} \textbf{\bibinfo{volume}{231}}, \bibinfo{pages}{826--840} (\bibinfo{year}{1979}).

\bibitem{koester2009}
\bibinfo{author}{{Koester}, D.}
\newblock \bibinfo{journal}{\bibinfo{title}{{Accretion and diffusion in white dwarfs. New diffusion timescales and applications to GD 362 and G 29-38}}}.
\newblock {\it {\JournalTitle{\aap}}} \textbf{\bibinfo{volume}{498}}, \bibinfo{pages}{517--525} (\bibinfo{year}{2009}).

\bibitem{tassoul1990}
\bibinfo{author}{{Tassoul}, M.}, \bibinfo{author}{{Fontaine}, G.} \& \bibinfo{author}{{Winget}, D.~E.}
\newblock \bibinfo{journal}{\bibinfo{title}{{Evolutionary Models for Pulsation Studies of White Dwarfs}}}.
\newblock {\it {\JournalTitle{\apjs}}} \textbf{\bibinfo{volume}{72}}, \bibinfo{pages}{335} (\bibinfo{year}{1990}).

\bibitem{cukanovaite2019}
\bibinfo{author}{{Cukanovaite}, E.} et~al.
\newblock \bibinfo{journal}{\bibinfo{title}{{Calibration of the mixing-length theory for structures of helium-dominated atmosphere white dwarfs}}}.
\newblock {\it {\JournalTitle{\mnras}}} \textbf{\bibinfo{volume}{490}}, \bibinfo{pages}{1010--1025} (\bibinfo{year}{2019}).

\bibitem{grossman1996}
\bibinfo{author}{{Grossman}, S.~A.} \& \bibinfo{author}{{Taam}, R.~E.}
\newblock \bibinfo{journal}{\bibinfo{title}{{Double-Diffusive Mixing-Length Theory, Semiconvection and Massive Star Evolution}}}.
\newblock {\it {\JournalTitle{\mnras}}} \textbf{\bibinfo{volume}{283}}, \bibinfo{pages}{1165--1178} (\bibinfo{year}{1996}).

\bibitem{paxton2018}
\bibinfo{author}{{Paxton}, B.} et~al.
\newblock \bibinfo{journal}{\bibinfo{title}{{Modules for Experiments in Stellar Astrophysics (MESA): Convective Boundaries, Element Diffusion, and Massive Star Explosions}}}.
\newblock {\it {\JournalTitle{\apjs}}} \textbf{\bibinfo{volume}{234}}, \bibinfo{pages}{34} (\bibinfo{year}{2018}).

\bibitem{wood2013}
\bibinfo{author}{{Wood}, T.~S.}, \bibinfo{author}{{Garaud}, P.} \& \bibinfo{author}{{Stellmach}, S.}
\newblock \bibinfo{journal}{\bibinfo{title}{{A New Model for Mixing by Double-diffusive Convection (Semi-convection). II. The Transport of Heat and Composition through Layers}}}.
\newblock {\it {\JournalTitle{\apj}}} \textbf{\bibinfo{volume}{768}}, \bibinfo{pages}{157} (\bibinfo{year}{2013}).

\bibitem{manseau2016}
\bibinfo{author}{{Manseau}, P.~M.}, \bibinfo{author}{{Bergeron}, P.} \& \bibinfo{author}{{Green}, E.~M.}
\newblock \bibinfo{journal}{\bibinfo{title}{{A Spectroscopic Search for Chemically Stratified White Dwarfs in the Sloan Digital Sky Survey}}}.
\newblock {\it {\JournalTitle{\apj}}} \textbf{\bibinfo{volume}{833}}, \bibinfo{pages}{127} (\bibinfo{year}{2016}).

\bibitem{bedard2025}
\bibinfo{author}{{B{\'e}dard}, A.} \& \bibinfo{author}{{Tremblay}, P.-E.}
\newblock \bibinfo{journal}{\bibinfo{title}{{The prototype double-faced white dwarf has a thin hydrogen layer across its entire surface}}}.
\newblock {\it {\JournalTitle{\mnras}}} \textbf{\bibinfo{volume}{538}}, \bibinfo{pages}{L69--L75} (\bibinfo{year}{2025}).

\bibitem{Chayer1995}
\bibinfo{author}{{Chayer}, P.}, \bibinfo{author}{{Fontaine}, G.} \& \bibinfo{author}{{Wesemael}, F.}
\newblock \bibinfo{journal}{\bibinfo{title}{{Radiative Levitation in Hot White Dwarfs: Equilibrium Theory}}}.
\newblock {\it {\JournalTitle{\apjs}}} \textbf{\bibinfo{volume}{99}}, \bibinfo{pages}{189} (\bibinfo{year}{1995}).

\bibitem{koester2014}
\bibinfo{author}{{Koester}, D.}, \bibinfo{author}{{G{\"a}nsicke}, B.~T.} \& \bibinfo{author}{{Farihi}, J.}
\newblock \bibinfo{journal}{\bibinfo{title}{{The frequency of planetary debris around young white dwarfs}}}.
\newblock {\it {\JournalTitle{\aap}}} \textbf{\bibinfo{volume}{566}}, \bibinfo{pages}{A34} (\bibinfo{year}{2014}).

\bibitem{Rouis2024}
\bibinfo{author}{{Ould Rouis}, L.~B.} et~al.
\newblock \bibinfo{journal}{\bibinfo{title}{{Constraints on Remnant Planetary Systems as a Function of Main-sequence Mass with HST/COS}}}.
\newblock {\it {\JournalTitle{\apj}}} \textbf{\bibinfo{volume}{976}}, \bibinfo{pages}{156} (\bibinfo{year}{2024}).

\bibitem{Dupuis1993}
\bibinfo{author}{{Dupuis}, J.}, \bibinfo{author}{{Fontaine}, G.}, \bibinfo{author}{{Pelletier}, C.} \& \bibinfo{author}{{Wesemael}, F.}
\newblock \bibinfo{journal}{\bibinfo{title}{{A Study of Metal Abundance Patterns in Cool White Dwarfs. II. Simulations of Accretion Episodes}}}.
\newblock {\it {\JournalTitle{\apjs}}} \textbf{\bibinfo{volume}{84}}, \bibinfo{pages}{73} (\bibinfo{year}{1993}).

\bibitem{wilson2021}
\bibinfo{author}{{Wilson}, D.~J.} et~al.
\newblock \bibinfo{journal}{\bibinfo{title}{{Discovery of a young pre-intermediate polar}}}.
\newblock {\it {\JournalTitle{\mnras}}} \textbf{\bibinfo{volume}{508}}, \bibinfo{pages}{561--574} (\bibinfo{year}{2021}).

\bibitem{panstarrs2020}
\bibinfo{author}{{Flewelling}, H.~A.} et~al.
\newblock \bibinfo{journal}{\bibinfo{title}{{The Pan-STARRS1 Database and Data Products}}}.
\newblock {\it {\JournalTitle{\apjs}}} \textbf{\bibinfo{volume}{251}}, \bibinfo{pages}{7} (\bibinfo{year}{2020}).

\bibitem{wright2010}
\bibinfo{author}{{Wright}, E.~L.} et~al.
\newblock \bibinfo{journal}{\bibinfo{title}{{The Wide-field Infrared Survey Explorer (WISE): Mission Description and Initial On-orbit Performance}}}.
\newblock {\it {\JournalTitle{\aj}}} \textbf{\bibinfo{volume}{140}}, \bibinfo{pages}{1868--1881} (\bibinfo{year}{2010}).

\bibitem{Bellm2019}
\bibinfo{author}{{Bellm}, E.~C.} et~al.
\newblock \bibinfo{journal}{\bibinfo{title}{{The Zwicky Transient Facility: System Overview, Performance, and First Results}}}.
\newblock {\it {\JournalTitle{\pasp}}} \textbf{\bibinfo{volume}{131}}, \bibinfo{pages}{018002} (\bibinfo{year}{2019}).

\bibitem{Ricker2015}
\bibinfo{author}{{Ricker}, G.~R.} et~al.
\newblock \bibinfo{journal}{\bibinfo{title}{{Transiting Exoplanet Survey Satellite (TESS)}}}.
\newblock {\it {\JournalTitle{Journal of Astronomical Telescopes, Instruments, and Systems}}} \textbf{\bibinfo{volume}{1}}, \bibinfo{pages}{014003} (\bibinfo{year}{2015}).

\bibitem{blouin2023}
\bibinfo{author}{{Blouin}, S.}, \bibinfo{author}{{B{\'e}dard}, A.} \& \bibinfo{author}{{Tremblay}, P.-E.}
\newblock \bibinfo{journal}{\bibinfo{title}{{Carbon dredge-up required to explain the Gaia white dwarf colour-magnitude bifurcation}}}.
\newblock {\it {\JournalTitle{\mnras}}} \textbf{\bibinfo{volume}{523}}, \bibinfo{pages}{3363--3375} (\bibinfo{year}{2023}).

\bibitem{althaus2005}
\bibinfo{author}{{Althaus}, L.~G.} et~al.
\newblock \bibinfo{journal}{\bibinfo{title}{{The formation and evolution of hydrogen-deficient post-AGB white dwarfs: The emerging chemical profile and the expectations for the PG 1159-DB-DQ evolutionary connection}}}.
\newblock {\it {\JournalTitle{\aap}}} \textbf{\bibinfo{volume}{435}}, \bibinfo{pages}{631--648} (\bibinfo{year}{2005}).

\bibitem{bedard2022}
\bibinfo{author}{{B{\'e}dard}, A.}, \bibinfo{author}{{Bergeron}, P.} \& \bibinfo{author}{{Brassard}, P.}
\newblock \bibinfo{journal}{\bibinfo{title}{{On the Spectral Evolution of Hot White Dwarf Stars. III. The PG 1159-DO-DB-DQ Evolutionary Channel Revisited}}}.
\newblock {\it {\JournalTitle{\apj}}} \textbf{\bibinfo{volume}{930}}, \bibinfo{pages}{8} (\bibinfo{year}{2022}).

\bibitem{werner2006}
\bibinfo{author}{{Werner}, K.} \& \bibinfo{author}{{Herwig}, F.}
\newblock \bibinfo{journal}{\bibinfo{title}{{The Elemental Abundances in Bare Planetary Nebula Central Stars and the Shell Burning in AGB Stars}}}.
\newblock {\it {\JournalTitle{\pasp}}} \textbf{\bibinfo{volume}{118}}, \bibinfo{pages}{183--204} (\bibinfo{year}{2006}).

\bibitem{Petit2005}
\bibinfo{author}{{Petitclerc}, N.}, \bibinfo{author}{{Wesemael}, F.}, \bibinfo{author}{{Kruk}, J.~W.}, \bibinfo{author}{{Chayer}, P.} \& \bibinfo{author}{{Bill{\`e}res}, M.}
\newblock \bibinfo{journal}{\bibinfo{title}{{FUSE Observations of DB White Dwarfs}}}.
\newblock {\it {\JournalTitle{\apj}}} \textbf{\bibinfo{volume}{624}}, \bibinfo{pages}{317--330} (\bibinfo{year}{2005}).

\bibitem{koester2014a}
\bibinfo{author}{{Koester}, D.}, \bibinfo{author}{{Provencal}, J.} \& \bibinfo{author}{{G{\"a}nsicke}, B.~T.}
\newblock \bibinfo{journal}{\bibinfo{title}{{Atmospheric parameters and carbon abundance for hot DB white dwarfs}}}.
\newblock {\it {\JournalTitle{\aap}}} \textbf{\bibinfo{volume}{568}}, \bibinfo{pages}{A118} (\bibinfo{year}{2014}).

\bibitem{bedard2024_review}
\bibinfo{author}{{B{\'e}dard}, A.}
\newblock \bibinfo{journal}{\bibinfo{title}{{The spectral evolution of white dwarfs: where do we stand?}}}
\newblock {\it {\JournalTitle{\apss}}} \textbf{\bibinfo{volume}{369}}, \bibinfo{pages}{43} (\bibinfo{year}{2024}).

\end{thebibliography}

\section*{Acknowledgements} 
We thank the two anonymous referees for their detailed and insightful reports which helped improve the quality of the paper. This project has received funding from the European Research Council (ERC) under the European Union’s Horizon 2020 research and innovation programme, grant agreements 101002408 (MOS100PC) and 101020057 (WDPLANETS). A.B. is a Postdoctoral Fellow of the Natural Sciences and Engineering Research Council (NSERC) of Canada. 
T.C. acknowledges support from NASA through the NASA Hubble Fellowship grant HST-HF2-51527.001-A awarded by the Space Telescope Science Institute, which is operated by the Association of Universities for Research in Astronomy, Inc., for NASA, under contract NAS5-26555.
This research is based on observations made with the NASA/ESA Hubble Space Telescope obtained from the Space Telescope Science Institute, which is operated by the Association of Universities for Research in Astronomy, Inc., under NASA contract NAS 5-26555, associated with program 15073. This work has made use of data from the European Space Agency (ESA) mission {\it Gaia} (\url{https://www.cosmos.esa.int/gaia}), processed by the {\it Gaia} Data Processing and Analysis Consortium (DPAC, \url{https://www.cosmos.esa.int/web/gaia/dpac/consortium}). Funding for the DPAC has been provided by national institutions, in particular, the institutions participating in the {\it Gaia} Multilateral Agreement. The authors acknowledge the University of Warwick's Scientific Computing Research Technology Platform (SCRTP) for assistance in the research described in this paper.
Observations reported here were obtained at the MMT Observatory, a joint facility of the Smithsonian Institution and the University of Arizona. This paper uses data products produced by the OIR Telescope Data Center (TDC), supported by the Smithsonian Astrophysical Observatory. The MMT/Binospec data were reduced by the SAO TDC; we thank Sean Moran and Igor Chilingarian for their help in that work.

\section*{Author contributions}
All authors contributed to the work presented in this paper. S.S. led the project and performed the spectral analysis. A.B. performed the envelope modelling. S.S. and A.B. wrote the major part of the manuscript. B.T.G. was the PI of the \textit{HST} program. B.T.G, P.-E.T., J.F., and J.J.H. helped in the interpretation of WD\,0525+526 and the writing of the manuscript. D.K. calculated the atmospheric models used in the spectral analysis. M.A.H. helped in the analysis of optical spectra. T.C. acquired the MMT/Binospec spectrum. S.R. helped in the interpretation of interstellar absorption in the spectra.

Correspondence and requests for materials should be addressed to Snehalata Sahu (snehalatash30@gmail.com) and Antoine B\'edard (antoine.bedard@warwick.ac.uk).

\section*{Ethics declarations}

The authors declare no competing interests.

\section*{Supplementary information}
\subsection*{Supplementary\,Data\,1} MMT/Binospec spectrum of WD\,0525+526. Columns are: air wavelength (\AA), flux (erg cm$^{-2}$ s$^{-1}$ \AA$^{-1}$), and flux error (erg cm$^{-2}$ s$^{-1}$ \AA$^{-1}$).

\subsection*{Supplementary\,Data\,2}
Best-fitting model spectrum for WD\,0525+526 (Fig.\,\ref{fig:hst_spec}) without convolution to COS spectral resolution. Columns are: vacuum wavelength (\AA), and model flux (erg cm$^{-2}$ s$^{-1}$ \AA$^{-1}$).
\subsection*{Supplementary\,Data\,3}
Model envelope chemical profile of WD\,0525+526 with helium at the upper limit (Fig.\,\ref{fig:env_model}). Columns are: logarithm of mass depth (logq), logarithm of hydrogen (logH), helium (logHe), and carbon (logC) mass fractions, and a number indicating the presence and type of convective mixing, where 0 denotes no mixing, 1 denotes normal convection, 2 denotes overshoot, and 3 denotes semi-convection.
\subsection*{Supplementary\,Data\,4} Model envelope chemical profile of WD\,0525+526 without any helium (Extended\,Data\,Fig.\,\ref{fig:env_model_nohe}). Columns are the same as for Supplementary\,Data\,3.
\subsection*{Supplementary\,Data\,5} Model envelope chemical profile of WD\,J2340$-$1819 with helium at the upper limit (Fig.\,\ref{fig:env_model}). Columns are the same as for Supplementary\,Data\,3.
\subsection*{Supplementary\,Data\,6} Model envelope chemical profile of WD\,J2340$-$1819 without any helium (Extended\,Data\,Fig.\,\ref{fig:env_model_nohe}). Columns are the same as for Supplementary\,Data\,3.

\begin{table*}
\caption{\justifying \textbf{Astrometric and spectroscopic parameters of WD\,0525+526.} The astrometric data are from Gaia\,DR3 at epoch J2016\cite{gaiadr3}. The COS spectroscopic parameters are based on a mass-radius relation considering a carbon-oxygen core and thin helium and hydrogen layers\cite{bedard2024}. The errors represent 1$\sigma$ statistical uncertainties in the measurements. Here, $v_{l}$ is the line of sight photospheric velocity not corrected for gravitational redshift and $v_{\rm ISM}$ is the line of sight velocity of the interstellar cloud. The upper limit on the helium abundance was obtained from the optical spectrum\cite{Gianninas2011}.} 
\label{table:params}
\centering 
\begin{tabular}{c c}
\hline\hline 
Parameter & value \\
\hline 
R.A. (J2016) [deg] & 82.46206 \\
Dec (J2016) [deg] & 52.66239\\
Gaia Source ID& 263082591016645504
\\
Parallax [mas]& $25.56\pm0.05$\\
Distance [pc] & $39.12\pm0.08$\\
$\mu_{\alpha}$ [mas yr$^{-1}$]& $+364.193\pm0.046$	\\
$\mu_{\delta}$ [mas yr$^{-1}$]& $-548.306\pm0.034$\\\hline
\Teff~[K]& $20\,820\pm96$\\
log ($g$ [cm\,s$^{-2}$])& $9.05\pm0.02$ \\
Radius [$\times10^{-3}$\Rsun] & $5.44\pm0.11$\\
Mass [\Msun]& $1.20\pm0.01$\\
$v_{l}$ [km s$^{-1}$] & $+92.8\pm0.6$\\
$v_{\rm ISM}$ [km s$^{-1}$]& $+18.7\pm1.1$\\
log\,(C/H) & $-4.62\pm0.04$\\
log\,(He/H) & $<-1.66$\\
\hline
\end{tabular}\\
%\footnotesize Here, $v_{l}$ is the line of sight photospheric velocity not corrected for gravitational redshift.\\ The upper limit in helium is obtained from optical spectra.
\end{table*}
%\section*{Extended Data}
%\section*{Figure Legends and Tables
\clearpage
\begin{extendfigure*}[t]
    \centerline{\includegraphics[width=0.45\hsize]{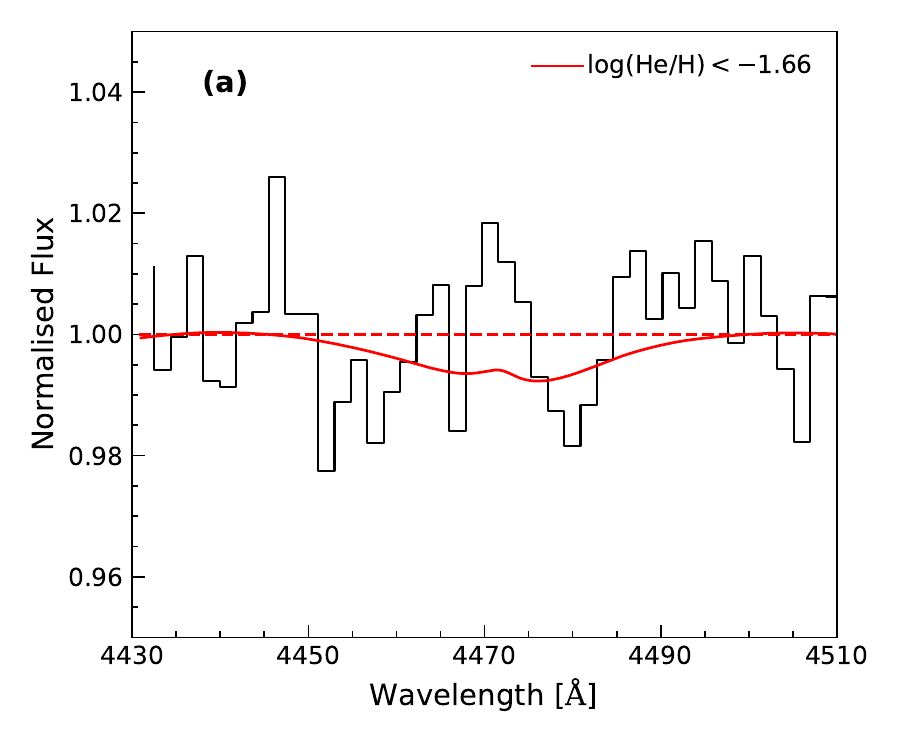}
\includegraphics[width=0.45\hsize]{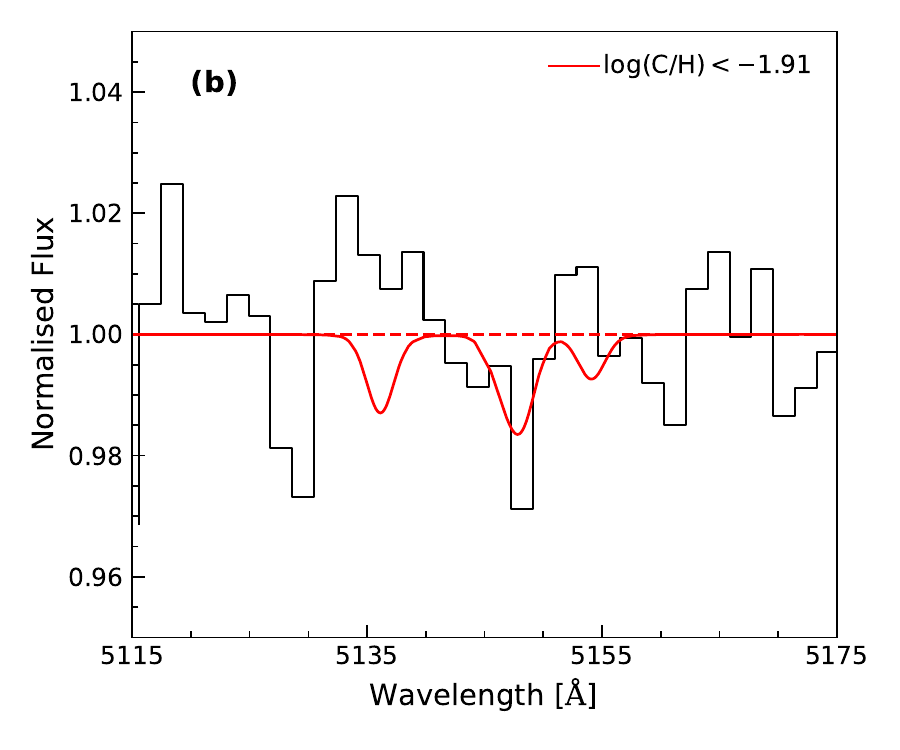}}
\caption{\justifying \textbf{Optical upper limits on the helium and carbon abundances of WD$\,$0525+526.} The left (a) and right (b) panels illustrate the upper limit determination for helium and carbon, respectively, using the optical spectroscopic data \cite{Gianninas2011} shown as solid black lines. The model spectra corresponding to the 99th percentile upper limits are over-plotted as solid red lines. To calculate the upper limits, we created a grid of models with log(He/H) varying from $-4.0$ to $-0.5$ and log(C/H) from $-8.0$ to $-1.0$ in steps of 0.05. For helium, we selected the spectral region 4430--4510\,\AA\ which contains a strong absorption feature from the \Ion{He}{i} line at 4472.75\,\AA. For carbon, we chose the spectral region spanning 5115--5175\,\AA\ with the strongest \Ion{C}{ii} lines from 5134.38\,\AA\ to 5152.52\,\AA. Fixing the \Teff, \logg, and velocity to the best-fit values obtained from ultraviolet spectroscopy, the upper limits were determined following the same procedure as for a previously identified DAQ white dwarf\cite{mark2020}. }
\label{fig:he_ul}
\end{extendfigure*}

% Right panel (b) shows the cumulative distribution function (CDF) calculated by integrating the likelihood (probability) function which is defined as the exponential of exp($-\chi^{2}_{r}/2$). The upper limit (red dashed line) corresponds to the 99\% confidence level of the CDF.}

\clearpage
\begin{extendfigure*}[t]
\centerline{\includegraphics[width=0.5\hsize]{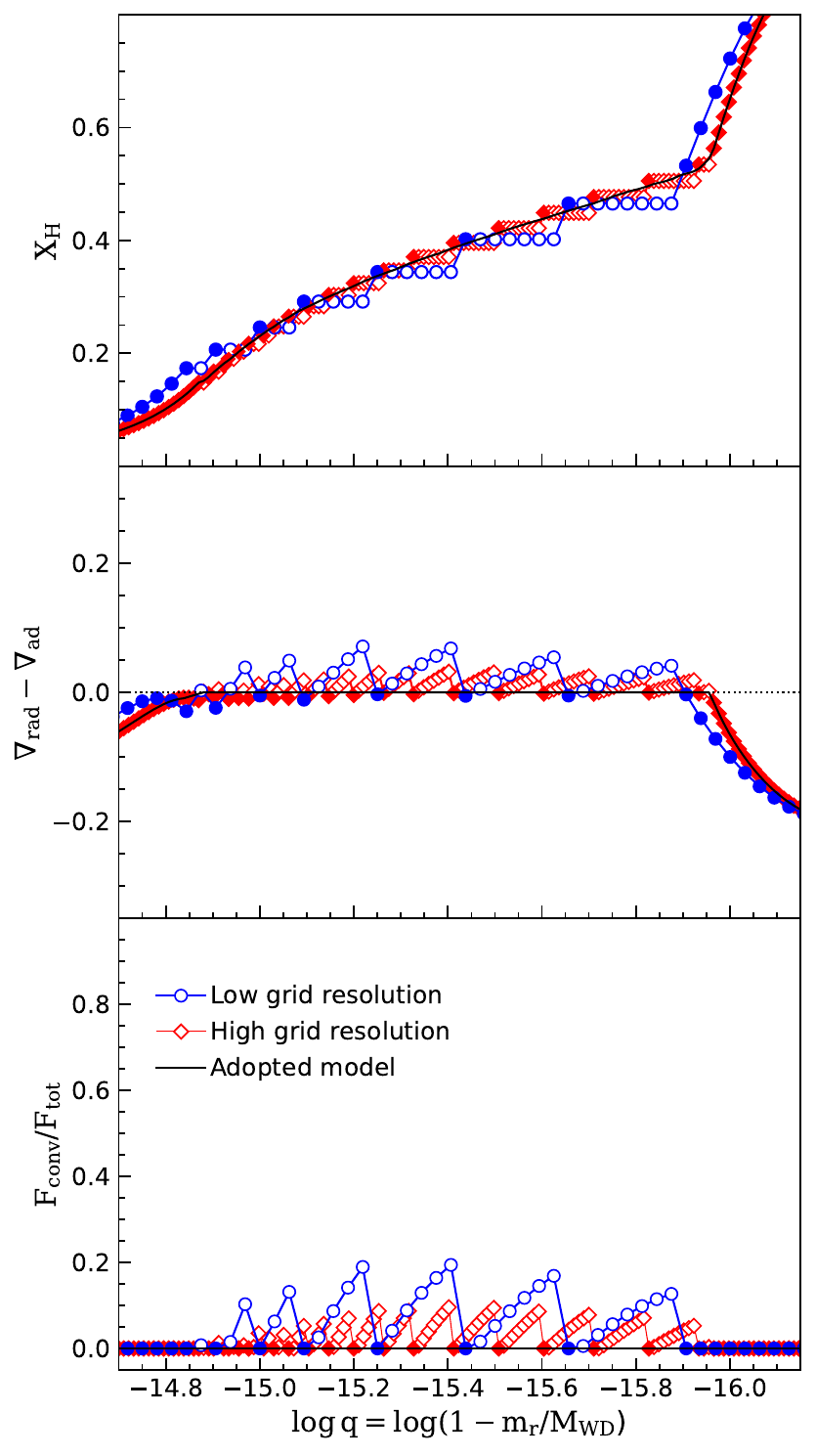}}
\caption{\justifying \textbf{Properties of the semi-convective region in envelope models with different grid resolutions.} The top, middle, and bottom panels show the hydrogen mass fraction, the difference between the radiative and adiabatic temperature gradients, and the ratio of convective to total energy flux, respectively, as a function of logarithmic mass depth. Illustrative models with low and high grid resolutions ($\simeq$35 and 100 grid points in the region of interest) are displayed as blue circles and red diamonds, respectively. Each grid point is depicted as an individual symbol, with empty and filled symbols denoting convective and non-convective grid points, respectively. The solid black lines show our adopted model, which was obtained by adjusting the chemical profile of the high-resolution model to enforce $\nabla_{\rm rad}$ = $\nabla_{\rm ad}$ in the semi-convective region. In the middle panel, the dotted black line indicates $\nabla_{\rm rad}$ = $\nabla_{\rm ad}$.}
\label{fig:semiconv1}
\end{extendfigure*}

\clearpage
\begin{extendfigure*}[t]
\centerline{\includegraphics[width=0.5\hsize]{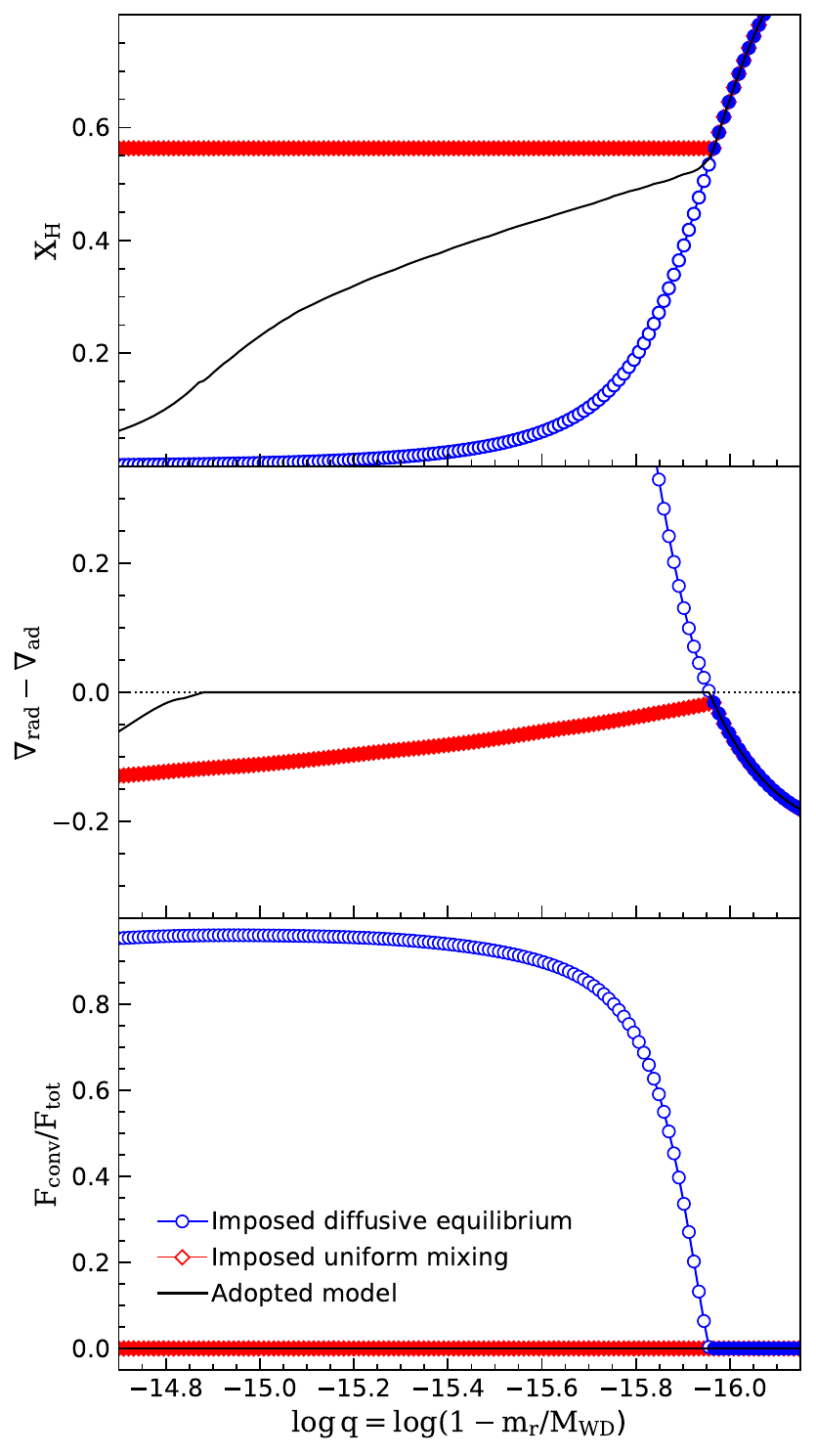}}
\caption{\justifying \textbf{Properties of the semi-convective region in envelope models with imposed chemical profiles.} he top, middle, and bottom panels show the hydrogen mass fraction, the difference between the radiative and adiabatic temperature gradients, and the ratio of convective to total energy flux, respectively, as a function of logarithmic mass depth. Illustrative high-resolution models with imposed diffusive equilibrium and imposed uniform mixing in the semi-convective region are displayed as blue circles and red diamonds, respectively. Each grid point is depicted as an individual symbol, with empty and filled symbols denoting convective and non-convective grid points, respectively. The model assuming diffusive equilibrium is convectively unstable, while the model assuming uniform mixing is convectively stable, therefore both models are inconsistent. The solid black lines show our adopted model, which is the only self-consistent solution.}
\label{fig:semiconv1_test}
\end{extendfigure*}

\clearpage
\begin{extendfigure*}[t]
\centerline{\includegraphics[width=0.8\hsize]{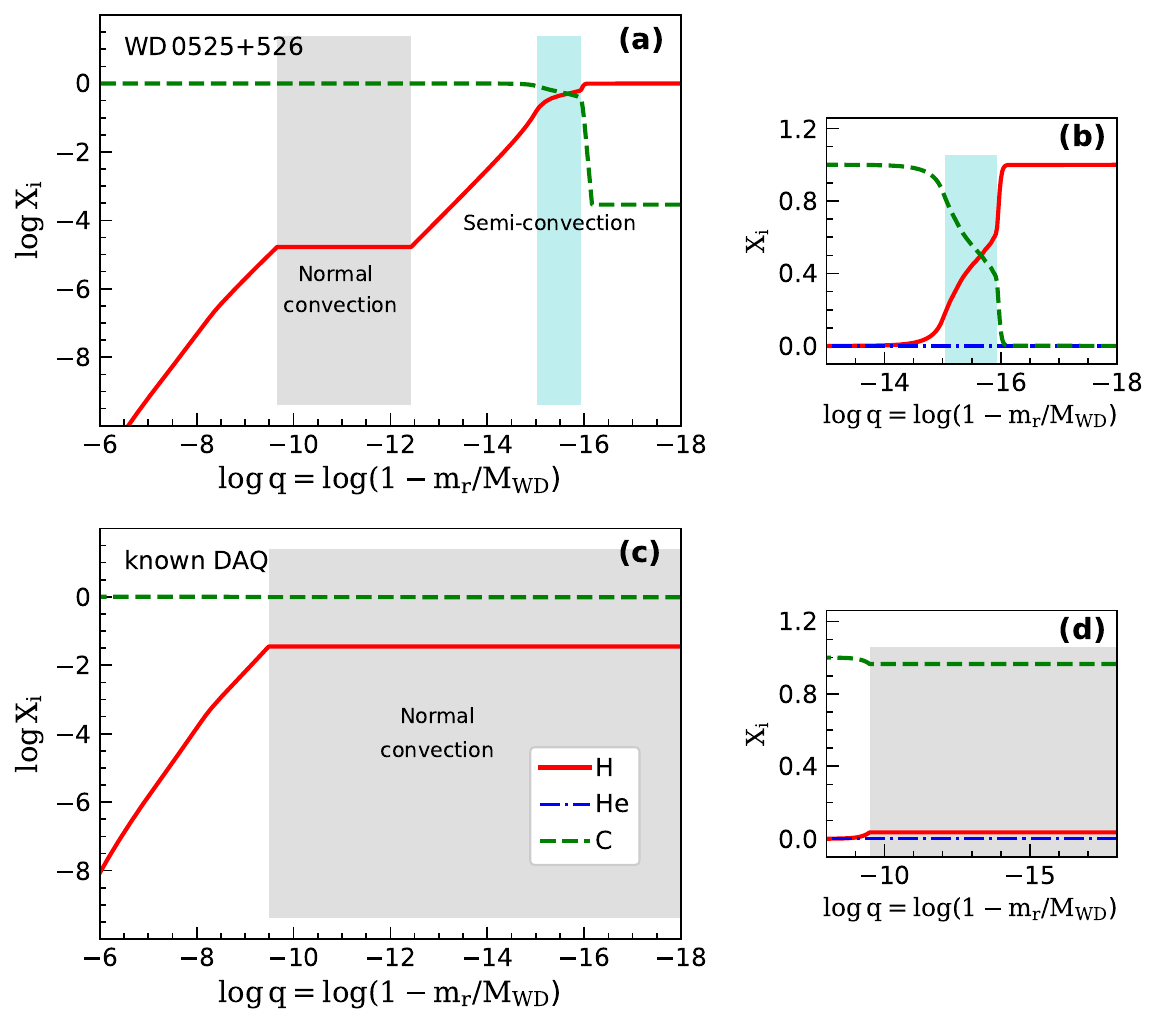}}
\caption{\justifying \textbf{Envelope chemical structure of DAQ white dwarfs without any helium.} The radial chemical profile of the model envelopes is expressed as the run of the elemental mass fractions with logarithmic mass depth, $\log q = \log (1-m_{r}/M_{\rm WD})$. The top panels (a and b) show the chemical profile of the hot DAQ WD$\,$0525+526 assuming the atmospheric parameters determined from ultraviolet spectroscopy (see Fig.\,\ref{fig:hst_spec}). The bottom panels (c and d) show the chemical profile of the cooler DAQ WD$\,$J2340$-$1819 assuming the atmospheric parameters\cite{kilic2024} $\Teff=15\,836\,$K, $\logg=8.95$, and log(C/H)$=+0.36$, which are representative of the six known DAQ white dwarfs. Normal convection zones are shaded in grey, while semi-convection zones are shaded in blue. These models do not include any photospheric helium.}
\label{fig:env_model_nohe}
\end{extendfigure*}

\end{document}